\documentclass{article}
\usepackage{amsmath}
\usepackage{amssymb}

\newcommand{\beq}{\begin{equation}}
\newcommand{\eeq}{\end{equation}}
\newcommand{\bea}{\begin{eqnarray}}
\newcommand{\eea}{\end{eqnarray}}

\begin{document}

\title{Hamiltonian Relativistic Two-Body Problem: Center of Mass and Orbit
Reconstruction }
\author{David Alba \\
Dipartimento di Fisica\\
Universita' di Firenze\\
Via G. Sansone 1\\
50019 Sesto Fiorentino (FI), Italy \thanks{%
alba@fi.infn.it} \and Horace W. Crater \\
The University of Tennessee Space Institute \\
Tullahoma, TN 37388 USA \thanks{%
hcrater@utsi.edu} \and Luca Lusanna \\
Sezione INFN di Firenze\\
Via G. Sansone 1\\
50019 Sesto Fiorentino (FI), Italy \thanks{%
lusanna@fi.infn.it}}
\maketitle

\begin{abstract}
After a short review of the history and problems of relativistic
Hamiltonian mechanics with action-at-a-distance inter-particle
potentials, we study isolated two-body systems in the rest-frame
instant form of dynamics. We give explicit expressions of the
relevant relativistic notions of center of mass, we determine the
generators of the Poincare' group in presence of interactions and
we show how to do the reconstruction of particles' orbits from the
relative motion and the canonical non-covariant center of mass. In
the case of a simple Coulomb-like potential model, it is possible
to integrate explicitly the relative motion and show the two
dynamical trajectories.
\end{abstract}

\section{Introduction}

In Newtonian mechanics the two-body problem is completely
understood both in configuration and phase space. The notions of
absolute time and absolute space allow us to describe the two
particles of mass $m_{i}$, $i=1,2$, with Euclidean position
3-vectors ${\vec{x}}_{i}$ and momenta ${\vec{p}}_{i}$ in an
inertial frame. For an isolated two-body system the Hamiltonian $
H = \sum_{i=1}^{2}\, {\frac{{{\vec{p}}_{i}^{2}}}{{2\,m_{i}}}} +
V(|{\vec{x}}_{1} - { \vec{x}}_{2}|)$ is the energy generator of
the kinematical Galilei group, whose other generators are all
interaction independent. With the point (both in coordinate and
momenta) canonical transformation ${\vec{x}} = {\frac{{ m_{1}\,
{\vec{x}}_{1} + m_{2}\, {\vec{x}}_{2}}}{{m_{1} + m_{2}}}}$,
$\vec{p} = {\vec{ p}}_{1} + {\vec{p}}_{2}$, $\vec{r} =
{\vec{x}}_{1} - {\vec{x}}_{2}$, $\vec{q} = { \frac{1}{2}}\,
({\vec{p}}_{1} - {\vec{p}}_{2})$ we can separate the decoupled
center of mass from the relative motion: the new Hamiltonian is $
H = H_{com} + H_{rel}$ with $H_{com} =
{\frac{{{\vec{p}}^{2}}}{{2\,m}}}$ ($ m = m_{1} + m_{2}$) and
$H_{rel} = {\frac{{{\vec{q}}^{2}}}{{2\,\mu }}} + V(|\vec r|)$
($\mu = { \frac{{m_{1}\, m_{2}}}{m}}$). The relative Hamiltonian
$H_{rel}$ governs the relative motion and, when its Hamilton
equations have been solved, the trajectories of the particles are
obtained with the inverse canonical transformation ${\vec{x}}_{1}
= \vec{x} + {\frac{{m_{2}}}{{m_{1} + m_{2}}}}\, \vec{r }$,
${\vec{x}}_{2} = \vec{x} - {\frac{{m_{1}}}{{m_{1} + m_{2}}}}\,
\vec{r}$, ${\vec{ p}}_{1} = {\frac{1}{2}}\, \vec{p} + \vec{q}$,
${\vec{p}}_{2} = {\frac{1}{2}}\, \vec{p } - \vec{q}$. As a
consequence the non-relativistic theory of orbits, for either 2 or
N particles, is well understood and developed (see for instance
Ref.\cite{1}). \bigskip

By contrast, in special relativity, where only Minkowski
space-time is absolute, where there is no absolute notion of
simultaneity and where inertial frames are connected by the
transformations generated by the kinematical Poincare' group, the
situation is extremely more complicated and till now there is no
completely self-consistent theory of orbits even for the two-body
case. This is due to the facts that\hfill\break

\noindent i) the particles locations and momenta are now 4-vectors
$x_{i}^{\mu }$, $p_{i}^{\mu }$;\hfill\break
 \noindent ii) the momenta are not independent but must satisfy mass-shell
conditions [since a relativistic particle is an irreducible
representation of the Poincare' group with mass $ m_{i}$ and a
value of the spin (only scalar particles will be studied in this
paper)]; \hfill\break
 \noindent iii) a simultaneity convention (for instance Einstein's one identifying
inertial frames) for the synchronization of distant clocks has to
be introduced, so that the time components $x_{i}^{o}$ are no more
independent;\hfill\break
 \noindent iv) the inter-particle interaction potentials appear in the boosts
as well as in the energy generator in the instant form of
dynamics;\hfill\break
 \noindent v) the structure of the Poincare' group implies that there is no
definition of relativistic 4-center of mass sharing all the
properties of the non-relativistic 3-center of mass.\medskip

Since a clarification of all these problems has recently been
obtained \cite {2}, in this paper we want to illustrate these
developments by using the simplest two-body system with a scalar
action-at-a-distance (a-a-a-d) interaction, for which a closed
Poincare' algebra can be found in the rest-frame instant form of
dynamics, as an example. By using the relativistic generalization
\cite{2} of the above quoted non-relativistic canonical basis, we
will show that the potential appearing in the energy Hamiltonian
(as with $H_{rel}$) determines the relative motion, while the
potentials appearing in the Lorentz boosts (which disappear in the
non-relativistic limit), together with the notion of the canonical
non-covariant 4-center of mass, contribute to the reconstruction
of the actual orbits of the two particles.\medskip

As a consequence for the first time we have full control on the relativistic
theory of orbits and we can start to reformulate at the relativistic level
the properties of the Newtonian theory of orbits . \bigskip

In Section II we give a brief history of the problems that have
arisen  in past attempts to develop Hamiltonian relativistic
mechanics. Then in Section III there is a review of the instant
form of dynamics, referred to above, with its two (external and
internal) realizations of the Poincare' algebra, while in Section
IV there is a review of the three intrinsic notions of
relativistic collective center-of-mass-like variables in both the
realizations. In Section V there is the relativistic extension of
the non-relativistic canonical transformation implementing the
separation of the center of mass from the relative variables and
how this can be used in general to do the reconstruction of the
particle orbits from the relative motion and the vanishing of the
internal canonical non-covariant 3-center of mass. In Section VI
there is the study of a simple two-body model with a-a-a-distance
interaction which correctly reproduces the Poincare' algebra
including potential-dependent boosts and energy generators, while
in Section VII there is the determination of its orbits with an
explicit integration of its equations of motion. A final
discussion on the relativistic theory of orbits with its avoidance
of the no-interaction theorem is given in the Conclusions. Finally
in Appendix A there is a review of the two-body models with first
class constraints.

\vfill\eject

\section{Brief History of Hamiltonian Relativistic Mechanics}

Relativistic classical particle mechanics with a-a-a-d
interactions and its Hamiltonian counterpart arose as an
approximation to interactions with a finite time delay (like the
electro-magnetic one) and have been quite useful in the treatment
of relativistic bound states with an instantaneous approximation
of the kernels of field-theoretic equations like the
Bethe-Salpeter equation. The starting points for Hamiltonian
relativistic particle mechanics were the instant, front and point
forms of relativistic Hamiltonian dynamics proposed by Dirac
\cite{3}. This approach was an attempt to find canonical
realizations of the Poincare' algebra such that some of the
generators, called Hamiltonians (the energy and the boosts in the
instant form), are not the direct sum of the corresponding ones
for free particles. After the pioneering works of Thomas,
Bakamjian and Foldy \cite{4}, employing $1/c$ expansions of the
Hamiltonians in the instant form (the only one analyzed in this
paper), many researches were done as can be seen from the
bibliography of Refs.\cite{5}.\medskip

The main obstacle in the development of models was the Currie -
Jordan - Sudarshan no-interaction theorem \cite{6}, whose
implication was the impossibility, in models with interaction, for
the canonical particle 4-positions to be 4-vectors when their time
components are put equal to the time of the reference inertial
frame. See Refs.\cite{7,8,9,10,11,12} for the localization problem
and Ref. \cite{13} for a review of the problem of the world-line
conditions and for the definition of the covariant particle
predictive 4-positions, coinciding with the canonical ones only in
absence of interactions.

From these studies it has become clear that relativistic particle
mechanics has to be formulated by using Dirac's theory of
constraints \cite{14} \footnote{ The symbol $\approx 0$(weakly
equal to zero) means that the constraint have been used to get the
equality. Let us remember that the constraints can be imposed only
after Poisson Brackets are evaluated.}: there must be as many
mass-shell first class constraints (containing the potentials of
the mutual interactions among the particles) as particles. The
first consistent two-body model with two first class constraints
depending upon a suitable potential was found by Droz-Vincent
\cite{15}, Todorov \cite{16} and Komar \cite{17} simultaneously
and independently (see Appendix A; for $N\geq 3$ a closed form of
the N first class constraints is not known).  These studies led to
the following problems: \medskip

a) The study of two- (and N-) particle configurations with a
one-to-one correlation among the world-lines. This can be done by
adding gauge fixing constraints, so that only the combination of
the original constraints describing the mass spectrum of the
global system of particles remains first class (moreover there are
$N-1$ pairs of second class constraints). Van Alstine \cite{18}
and the authors of Refs.\cite{19} developed  consistent two-body
models with  second class constraints. The avoidance of the
relative times in these models has been recently re-interpreted in
Ref.\cite{20} as the problem of the synchronization of the clocks
associated to the individual particles.
\medskip

b) The reformulation of the N-body problem with N first class
constraints as a form of N-times dynamics \cite{15,21}. As shown
in Ref.\cite{13}, in this way one can write the equations defining
Droz-Vincent covariant non-canonical predictive 4-coordinates
\cite{15}: each of them depends only on the proper time of the
associated particle, while the canonical coordinates depend
simultaneously on all the proper times when interactions are
present (or equivalently on the proper time of the center of mass
of the isolated system in the one-time theory). A by-product of
these studies was the reformulation of the Newtonian N-body
problem as a N-times theory \cite{22}: as a consequence a form of
no-interaction theorem appears also at the non-relativistic level.

\medskip

c) The identification of canonical bases containing a relativistic
4-center of mass and relativistic relative variables starting from
the original canonical 4-vectors $x_{i}^{\mu }$, $p_{i\mu }$. This
was a highly non-trivial problem due to the lack of a unique
notion of relativistic center of mass. If we use only the
Poincare' generators of the N-particle system, it is possible to
define only three such notions: a canonical non-covariant
Newton-Wigner-like 3-center of mass \cite{7,10,11}, a
non-canonical non-covariant M$\o $ller 3-center of energy \cite{8}
and a non-canonical covariant Fokker-Pryce 3-center of inertia
\cite{9,10}. \ Each then have to be extended to 4-centers
(${\tilde x}^{\mu}$, $R^{\mu}$ and $Y^{\mu}$, respectively). The
two non-covariant 4-centers ${\tilde x}^{\mu}$ and $R^{\mu}$
describe frame-dependent pseudo-world-lines filling the so-called
M$\o $ller world-tube \cite{8,21,23} around the world-line of the
Fokker-Pryce 4-center of inertia $Y^{\mu}$ (in the rest frame the
3 centers coincide). The invariant M$\o $ ller radius of this
world-tube is determined by the Poincare' Casimirs of the particle
configuration, $\rho =S/Mc$.

In Ref.\cite{2} there is a full classification of these centers
and of their properties and a methodology to find canonical bases
of center-of-mass and relative variables, also in the presence of
interactions, in the framework of the \textit{Wigner-covariant
rest-frame instant form of dynamics} developed in
Refs.\cite{21,23}. This instant form is a special case of
\textit{parametrized Minkowski theories} \cite{21,23} \footnote{
This approach was developed to give a formulation of the N-body
problem on arbitrary simultaneity 3-surfaces (corresponding to a
convention for the synchronization of distant clocks \cite{20}),
which could be also used as Cauchy surfaces for the
electro-magnetic field if present. In this way both the unknown
closed form of first class constraints and the special choices of
gauge fixings leading to second class ones are avoided. Moreover,
the change of clock synchronization convention may be formulated
as a \textit{ gauge transformation} not altering the physics and
there is no problem in introducing the electro-magnetic field when
the particles are charged. The rest-frame instant form corresponds
to the gauge choice of the 3+1 splitting whose simultaneity
3-surfaces are the intrinsic rest frame of the given configuration
of the isolated system.}, in which the leaves of the 3+1 splitting
of Minkowski space-time are inertial hyper-planes (simultaneity
3-surfaces called \textit{Wigner hyper-planes}) orthogonal to a
4-vector $P^{\mu}$, coinciding with the conserved 4-momentum
$P^{\mu }_{sys}$ of the N-particle system in the rest frame: if we
define the invariant mass $M = \sqrt{P^2_{sys}}$, then we have
$P^{\mu} = M\, u^{\mu}(P)$, $u^2(P) = 1$. The 4-vector $P^{\mu}$
is canonically conjugate to the canonical non-covariant 4-center
of mass ${\tilde x}^{\mu}$. Therefore the Wigner hyperplane at
time $\tau$ is the intrinsic rest frame of the isolated system at
time $\tau$.

In this approach all the particles depend on the scalar rest-frame
time $\tau = Y^{\mu}\, u_{\mu}(P) = {\tilde x}^{\mu}\, u_{\mu}(P)
= R^{\mu}\, u_{\mu}(P)$,  the Fokker-Pryce 4-center of inertia
$Y^{\mu }(\tau )$ is the inertial observer origin of the inertial
rest frame, and all the first class constraints have been solved
to determine the single particle energies. As a consequence we
have:

\noindent i) $x^{\mu}_i(\tau ) = Y^{\mu}(\tau ) +
\epsilon^{\mu}_r(P)\, \eta^r_i(\tau )$, $i=1,..,N$, where
$\epsilon^{\mu}_r(P)$ are suitable momentum-dependent space-like
4-vectors orthogonal to $u^{\mu}(P)$;

\noindent ii) $p^{\mu}_i(\tau )$ are suitable solutions of the
mass shell constraints. Therefore the particles must have a
definite sign of the energy (assumed positive in this paper) and
their independent canonical variables are the (Wigner spin-1)
position 3-vectors ${\vec \eta}_i(\tau )$ and their conjugate
3-momenta ${\vec \kappa}_i(\tau )$.

\noindent iii) There is an \textit{external} realization of the
Poincare' algebra with generators $P^{\mu }$, $J^{\mu \nu }$
(describing the properties of the Wigner hyper-planes in every
inertial frame) and an \textit{unfaithful internal} (called
unfaithful since some generators are weakly vanishing) realization
of the Poincare' algebra, with generators $M$ (the invariant mass
of the system), $ \vec{p}$, $\vec{j}$, $\vec{k}$ \footnote{ The
internal total 3-momentum is weakly vanishing, $\vec{p}\approx 0$,
since these three first class constraints define the rest frame.
If the internal Lorentz boosts $\vec{k}$ are put weakly equal to
zero, $\vec{k}\approx 0$, as gauge fixing constraints to the
rest-frame conditions, then it can be shown \cite{2,23} that the
3-position describing the collective center variable inside the
Wigner hyper-planes (the 3 internal centers coincide due to the
rest-frame conditions as shown in Section IV) coincides with the
origin, i.e. is located at the position of the external
Fokker-Pryce 4-center of inertia at each instant. Therefore only
the external canonical non-covariant 4-center of mass remains as a
decoupled point particle, without any double counting.} describing
the covariance properties inside the Wigner hyper-planes. The
Hamiltonian for the relative motion on the Wigner hyper-plane
(replacing the non-relativistic $H_{rel}$) is the the internal
energy generator $M$.

\noindent iv) It is now clear that the avoidance of the
no-interaction theorem implies the non-covariance of the canonical
external Newton-Wigner-like 4-center of mass ${\tilde x}^{\mu}$ (a
pseudo-world-line intersecting each Wigner hyper-plane in every
inertial frame and coinciding with the external Fokker-Pryce
center of inertia only when the reference inertial frame coincides
with the rest frame): this non-covariance is universally (i.e.
independently from the relativistic isolated system under
investigation) concentrated on the point-particle-like degrees of
freedom describing the \textit{decoupled} external center of mass
of the system.

\medskip

d) Another problem is the identification of special models suited
to the relativistic bound state problem. The model building
\cite{13,21,24,25,26} initially concentrated on the potentials in
the energy Hamiltonian, which governs the relative motion (the
canonical non-covariant 4-center of mass has free motion). The
much harder problem to find the suitable potentials in the Lorentz
boosts \cite{5}, so that the global Poincare' algebra is
satisfied, was finally solved in Ref.\cite{27} for charged scalar
particles interacting with a dynamical electro-magnetic field
(with Grassmann-valued electric charges to regularize the
self-energies): in the sector of configurations without an
independent radiation field the Darwin potential appeared in the
energy Hamiltonian (till now it had been obtained only coming down
from quantum field theory through instantaneous approximations to
the Bethe-Salpeter equation) and suitable related potentials in
the boost Hamiltonians. In Ref.\cite{28} analogous results
(involving the Salpeter potential) were obtained for charged
spinning particles (with Grassmann-valued spins implying Dirac
spin 1/2 fermions after quantization).

\vfill\eject

\section{The Inertial Rest-Frame Instant Form of Dynamics}

In this Section we will describe the main properties of the
inertial rest-frame instant form of dynamics in the case of free
particles by anticipating the various notions of relativistic
center of mass, which will be clarified in Section IV. In Appendix
B there is the definition of non-inertial rest frames, obtained as
other special gauges of parametrized Minkowski theories. \bigskip

As said in the previous Section, in the rest-frame instant form of
dynamics Minkowski space-time \footnote{ We use the metric $\eta
_{\mu \nu }=\,(+---).$} is foliated with inertial hyper-planes
(named Wigner hyper-planes) orthogonal to a 4-vector $P^{\mu}$
coinciding with the conserved 4-momentum $P_{sys}^{\mu }$ of the
given isolated system in the rest frame. With respect to an
arbitrary inertial frame the Wigner hyper-planes are described by
the following embedding

\begin{equation}
z^{\mu }(\tau ,\vec{\sigma}) = x_{s}^{\mu }(\tau ) + \epsilon
_{r}^{\mu }(u(P))\, \sigma ^{r},
  \label{1}
\end{equation}

\noindent with $x_{s}^{\mu }(\tau )$ being the world-line of an
arbitrary inertial observer. The (so-called \textit{radar}
\cite{20}) observer-dependent 4-coordinates $(\tau ;\vec{\sigma})$
are the proper time $\tau $ of this observer and 3-coordinates on
the Wigner hyper-planes $\Sigma _{\tau }$ having the observer as
origin $\vec{\sigma}=0$ for every $\tau $. The space-like
4-vectors $ \epsilon _{r}^{\mu }(u(P))$ together with the
time-like one $\epsilon _{o}^{\mu }(u(P))$ are the columns of the
standard Wigner boost for time-like Poincare' orbits \footnote{ It
sends the time-like four-vector $P^{\mu }$ to its rest-frame form
$ \overset{\circ }{P}{}^{\mu } = \eta \,\sqrt{P^{2}}\,
(1;\vec{0})$, where $\eta = sign\, P^{o}$; see Refs.\cite{2,21}.
From now on we restrict ourselves to positive energies, i.e. $\eta
=1$. While $\epsilon _{o}^{\mu }(u(P))$ and $ \epsilon _{\mu
}^{o}(u(P))$ are 4-vectors, $\epsilon _{r}^{\mu }(u(P))$ have more
complex transformation properties under Lorentz transformations,
given in the Conclusions.}

\begin{eqnarray}
\epsilon _{o}^{\mu }(u(P)) &=& u^{\mu }(P) = P^{\mu
}/\sqrt{P^{2}},\qquad \epsilon _{r}^{\mu }(u(P)) = (- u_{r}(P);
\delta_{r}^{i} - {\frac{{u^{i}(P)\, u_{r}(P)}}{{1 + u^{o}(P)}}}),  \notag \\
&&{}  \notag \\
\epsilon _{\mu }^{o}(u(P)) &=& \eta ^{o\alpha}\, \eta _{\mu \nu
}\, \epsilon _{\alpha}^{\nu }(u(P)) = u_{\mu }(P),\qquad \epsilon
_{\mu }^{r}(u(P)) = \eta ^{r\alpha}\, \eta _{\mu \nu }\, \epsilon
_{\alpha}^{\nu}(u(P)).\nonumber \\
 &&{}
  \label{2}
\end{eqnarray}

Since we are in the rest frame, we have $\tau \equiv T_s = u(P) \cdot x_s$.
Here $T_s$ is the scalar rest time of the inertial observer, whose
world-line is given by

\begin{equation}
x^{\mu}_s(\tau ) = x^{\mu}(0) + u^{\mu}(P)\, \tau .
 \label{3}
\end{equation}

In the rest-frame instant form the particles' 4-coordinates,
describing their world-lines, and the associated momenta (see
Refs.\cite{2,21}) are

\begin{eqnarray}
x_{i}^{\mu }(\tau ) &=& z^{\mu }(\tau ,{\vec{\eta}}_{i}(\tau )) =
x_{s}^{\mu}(\tau ) + \epsilon _{r}^{\mu }(u(P))\, \eta _{i}^{r}(\tau ),
 \notag \\
p_{i}^{\mu }(\tau ) &=& \eta _{i}\, \sqrt{m_{i}^{2} +
{\vec{\kappa}} _{i}^{2}(\tau )}\, u^{\mu }(P) + \epsilon _{r}^{\mu
}(u(P))\, \kappa _{ir}(\tau )\,\,\Rightarrow p_{i}^{2} =
m_{i}^{2},\nonumber \\
 &&{}\nonumber \\
 \Rightarrow && P_{sys}^{\mu} = \sum_{i=1}^N\, p_i^{\mu}
= u^{\mu}(P)\, \sum^N_{i=1}\, \eta_i\, \sqrt{m_{i}^{2} +
{\vec{\kappa}} _{i}^{2}(\tau )} + \epsilon^{\mu}_r(u(P))\,
\sum^N_{i=1}\, \kappa_{ir}(\tau ).
  \label{4}
\end{eqnarray}

This shows that inside the Wigner hyper-planes the N particles
(all assumed to have positive energy, i.e. $\eta _{i}=1$) are
described by the 6N Wigner spin-1 3-vectors ${\vec{\eta}}_{i}(\tau
)$, ${\vec{\kappa}}_{i}(\tau )$ \footnote{Under Lorentz
transformations $\Lambda $ these 3-vectors rotate with Wigner
rotations [$\overset{\circ }{P} = M\, (1;\vec{0})$; $P^{\mu } =
L(P, \overset{ \circ }{P})^{\mu }{}_{\nu }\, \overset{\circ
}{P}^{\nu } = \epsilon _{A = \nu }^{\mu }(u(P))\, \overset{\circ
}{P}^{\nu }$ with $L$ the standard Wigner boost]\hfill \break
\begin{eqnarray*}
R^{\mu }{}_{\nu }(\Lambda ,P) &=& {[L(\overset{\circ }{P},P)\,
\Lambda ^{-1}\, L(\Lambda P, \overset{\circ }{P})]}^{\mu }{}_{\nu
} = \left(
\begin{array}{cc}
1 & 0 \\
0 & R^{i}{}_{j}(\Lambda ,P)
\end{array}
\right) , \\
{} &&{} \\
R^{i}{}_{j}(\Lambda ,P) &=& {(\Lambda ^{-1})}^{i}{}_{j} -
{\frac{{(\Lambda ^{-1})^{i}{}_{o}\, P_{\beta }\, (\Lambda
^{-1})^{\beta }{}_{j}}}{{P_{\rho
}\, (\Lambda ^{-1})^{\rho }{}_{o} + \sqrt{P^{2}}}}} - \\
&-& {\frac{{P^{i}}}{{P^{o} + \sqrt{P^{2}}}}}\, [(\Lambda
^{-1})^{o}{}_{j} - {\frac{{ ((\Lambda ^{-1})^{o}{}_{o} - 1)\,
P_{\beta }\, (\Lambda ^{-1})^{\beta }{}_{j}}}{{ P_{\rho }\,
(\Lambda ^{-1})^{\rho }{}_{o} + \sqrt{P^{2}}}}}].
\end{eqnarray*}
As a consequence the scalar product of two of these 3-vectors is a
Lorentz scalar.} as independent canonical variables [$\{ \eta
_{i}^{r}(\tau ), \kappa _{js}(\tau )\} = \delta _{ij}\, \delta
_{s}^{r}$, $\{\eta _{i}^{r}(\tau ), \eta _{j}^{s}(\tau )\} =
\{\kappa _{ir}(\tau ), \kappa _{js}(\tau )\}=0$]. To them we must
add $P^{\mu } = \sum_i\, p_i^{\mu }$ and a canonically conjugate
collective variable ${\tilde{x}}^{\mu }$ [$\{ x^{\mu}_s, P^{\nu}
\} = \{ {\tilde x}^{\mu}, P^{\nu} \} = - \eta^{\mu\nu}$].

It turns out \cite{2} (see also Section IV) that the relevant
collective variable to be added is the \textit{external} canonical
non-covariant 4-center of mass ($M = \sqrt{\,P^{2}}$) \footnote{
As shown in Ref.\cite{21}, the three spin tensors $S^{\mu\nu} =
\epsilon^{\mu}_A(u(P))\, \epsilon^{\nu}_B(u(P))\, {\bar S}^{AB}$,
${\tilde S}^{\mu\nu}$ and ${\bar S}^{AB}$ satisfy the Lorentz
algebra: $\{ S^{\mu\nu}, S^{\alpha\beta} \} =
C^{\mu\nu\alpha\beta}_{\gamma\delta}\, S^{\gamma\delta}$, $\{
{\tilde S}^{\mu\nu}, {\tilde S}^{\alpha\beta} \} =
C^{\mu\nu\alpha\beta}_{\gamma\delta}\, {\tilde S}^{\gamma\delta}$,
$\{ {\bar S}^{AB}, {\bar S}^{CD} \} = C^{ABCD}_{EF}\, {\bar
S}^{EF}$, where $C^{\mu\nu\alpha\beta}_{\gamma\delta} =
\eta^{\nu}_{\gamma}\, \eta^{\alpha}_{\delta}\, \eta^{\mu\beta} +
\eta^{\mu}_{\gamma}\, \eta^{\beta}_{\delta}\, \eta^{\nu\alpha} -
\eta^{\nu}_{\gamma}\, \eta^{\beta}_{\delta}\, \eta^{\mu\alpha} -
\eta^{\mu}_{\gamma}\, \eta^{\alpha}_{\delta}\, \eta^{\nu\beta}$,
$C^{ABCD}_{EF} = \eta^{B}_{E}\, \eta^{C}_{F}\, \eta^{AD} +
\eta^{A}_{E}\, \eta^{D}_{F}\, \eta^{BC} - \eta^{B}_{E}\,
\eta^{D}_{F}\, \eta^{AC} - \eta^{A}_{E}\, \eta^{C}_{F}\,
\eta^{BD}$ are the Lorentz structure constants.}

\begin{eqnarray}
{\tilde{x}}^{\mu }(\tau ) &=& ({\tilde{x}}^{o}(\tau );
{\vec{\tilde{x}}}(\tau )) = z^{\mu }(\tau ,{\vec{\tilde{\sigma}}}) =  \notag \\
&=& x_{s}^{\mu }(\tau ) - {\frac{1}{{M\, (P^{o} + M)}}}\,
\Big[P_{\nu }\, S^{\nu \mu } + M\, (S^{o\mu } - S^{o\nu }\,
{\frac{{P_{\nu }\, P^{\mu }}}{{M^{2}}}})\Big],
\nonumber \\
&&  \notag \\
S^{\mu \nu } &=& J^{\mu\nu} - (x^{\mu}_s\, P^{\nu} - x^{\nu}_s\, P^{\mu})
=\nonumber \\
 &=& [u^{\mu }(P)\, \epsilon _{r}^{\nu }(u(P)) - u^{\nu
}(P)\, \epsilon _{r}^{\mu }(u(P))]\, {\bar{S}}^{or} + \epsilon
_{r}^{\mu}(u(P))\, \epsilon _{s}^{\nu }(u(P))\, {\bar{S}}^{rs},  \notag \\
{\bar{S}}^{rs} &= & \sum_{i=1}^{N}\, (\eta _{i}^{r}\, \kappa
_{i}^{s} - \eta _{i}^{s}\, \kappa _{i}^{r}),\quad {\bar{S}}^{or} =
- \sum_{i=1}^{N}\, \eta_{i}^{r}\, \sqrt{m_{i}^{2}\, +
{\vec{\kappa}}_{i}^{2}},  \notag \\
 {\tilde S}^{\mu\nu} &=& J^{\mu\nu} - ({\tilde x}^{\mu}\, P^{\nu} -
{\tilde x}^{\nu}\, P^{\mu}),\qquad {\tilde{S}}^{ij} = \delta
^{ir}\, \delta ^{js}\, {\bar{S}}^{rs},\qquad  { \tilde{S}}^{oi} =
- {\frac{{\delta ^{ir}\, {\bar{S}}^{rs}\, P^{s}}}{{P^{o} + M}}}.
\nonumber \\
 &&{}
 \label{5}
\end{eqnarray}

The point with coordinates ${\tilde{x}}^{\mu }(\tau )$ is the
decoupled canonical \textit{external 4-center of mass}, playing
the role of a kinematical external 4-center of mass and of a
decoupled observer with his parametrized clock (\textit{point
particle clock}). As shown in Refs.\cite {21}, \cite{2}, when we
restrict the parametrized Minkowski theory to the rest-frame
instant form and we add the gauge fixing $\vec{k}\approx 0$ (see
footnote 3), we get the result ${\dot{x}}_{s}^{\mu }(\tau ) =
{\dot{\tilde{x}}} ^{\mu }(\tau ) = u^{\mu }(P)$. Therefore the
gauge $\vec{k}\approx 0$ is the \textit{natural} one, because only
in it are both the velocities ${\dot{x}} _{s}^{\mu }(\tau )$ and
${\dot{\tilde{x}}}^{\mu }(\tau )$ parallel to $ P^{\mu }$, so that
there is no \textit{classical zitterbewegung} in the associated
world-lines.

\bigskip

In conclusion, as a consequence of Eqs.(\ref{5}), in the
rest-frame instant form the standard $8N$ variables $x_{i}^{\mu
}$, $p_{i}^{\mu }$ are re-expressed in terms of the 8 variables
$x_{s}^{\mu }$ (or ${\tilde{x}} ^{\mu }$), $P^{\mu } = M\, u^{\mu
}(P)$ and the $6N$ variables ${\vec{\eta}} _{i} $,
${\vec{\kappa}}_{i}$ restricted by the rest-frame condition
$\vec{p} \approx 0$ and by the associated gauge-fixings (see
footnote 3). Therefore we have $8 + 6N - 6 = 2\, (3N + 1)$
variables: the lacking $2\, (N - 1)$ to arrive to $8N$ are the
$N-1$ relative times (they disappeared due to the clock
synchronization) and $N - 1$ relative momenta (they disappeared
because, as shown in Eqs.(\ref{4}), the particles are on the mass
shell).

\bigskip

Let us stress that the rest-frame instant form succeeds in
separating the relativistic center of mass from the relative
motion by means of a \textit{ splitting} of the description of the
isolated system into an \textit{external } one and an
\textit{internal} one.

\medskip

A) The external description is concerned with the embedding of the
Wigner hyper-planes into Minkowski space-time from the point of
view of a generic inertial observer. Each Wigner hyper-plane,
orthogonal to the conserved 4-momentum $P^{\mu}$ of the isolated
system, is parametrized by means of canonical coordinate ${\tilde
x}^{\mu}$ (conjugate to $P^{\mu}$), which describes the decoupled
collective degrees of freedom of the isolated system (at the
non-relativistic level it is the free center of mass $\vec x$ with
Hamiltonian $H_{com} = {\frac{{{\vec p}^2}}{{2m}}}$). There is an
\textit{ external} realization of the Poincare' algebra, which
governs the covariance properties of Wigner hyper-planes under
Poincare' transformations $(\Lambda , a)$.

It can be shown \cite{2} that its generators have the following
form [$M = \sqrt{\,P^{2}}$; while $i,j..$ are Euclidean indices,
$r,s..$ are Wigner spin-1 indices; ${\tilde S}^{\mu\nu}$ is given
in Eq.(\ref{5}); footnote 7 implies $\{ P^{\mu}, P^{\nu} \} = 0$,
$\{ P^{\mu}, J^{\alpha\beta} \} = \eta^{\mu\alpha}\, P^{\beta} -
\eta^{\mu\beta}\, P^{\alpha}$, $\{ J^{\mu\nu}, J^{\alpha\beta} \}
= C^{\mu\nu\alpha\beta}_{\gamma\delta}\, J^{\gamma\delta}$]

\begin{eqnarray}
P^{\mu } &,&\qquad J^{\mu \nu } = x_{s}^{\mu }\, P^{\nu } -
x_{s}^{\nu }\, P^{\mu } + S^{\mu \nu } = {\tilde{x}}^{\mu }\,
P^{\nu } - {\tilde{x}}^{\nu }\, P^{\mu } + {
\tilde{S}}^{\mu \nu },  \notag \\
&&{}  \notag \\
P^{o} &=& \sqrt{M^{2} + {\vec{P}}^{2}},  \notag \\
&&{}  \notag \\
J^{ij} &=& {\tilde{x}}^{i}\, P^{j} - {\tilde{x}}^{j}\, P^{i} +
\delta ^{ir}\, \delta ^{js}\, \sum_{i=1}^{N}\, (\eta _{i}^{r}\,
\kappa_{i}^{s} - \eta _{i}^{s}\, \kappa _{i}^{r}) =\nonumber \\
 &=& {\tilde{x}}^{i}\, P^{j} - {\tilde{x}}^{j}\, P^{i} + \delta ^{ir}\, \delta
^{js}\, \epsilon ^{rsu}\, {\bar{S}}^{u},  \notag \\
K^{i} &=& J^{oi} = {\tilde{x}}^{o}\, P^{i} - {\tilde{x}}^{i}\,
\sqrt{M^{2} + {\vec{P}}^{2}} - \nonumber \\
 &-& {\frac{1}{{M + \sqrt{M^{2} + {\vec{P}}^{2}}}}}\, \delta
^{ir}\, P^{s}\, \sum_{i=1}^{N}\, (\eta _{i}^{r}\, \kappa _{i}^{s}
- \eta_{i}^{s}\, \kappa _{i}^{r}) =  \notag \\
&=& {\tilde{x}}^{o}\, P^{i} - {\tilde{x}}^{i}\, \sqrt{M^{2} +
{\vec{P}}^{2}} - {\frac{ {\delta ^{ir}\, P^{s}\, \epsilon ^{rsu}\,
{\bar{S}}^{u}}}{{M + \sqrt{M^{2} + {\vec{P }}^{2}}}}}.
  \label{6}
\end{eqnarray}

Note that both ${\tilde L}^{\mu\nu} = {\tilde x}^{\mu}\, P^{\nu} -
{\tilde x} ^{\nu}\, P^{\mu}$ and ${\tilde S}^{\mu\nu}$ are
conserved.\medskip \medskip

It is this external realization which implements the Wigner
rotations of footnote 6. \medskip

Let us remark that this realization is universal in the sense that
it depends on the nature of the isolated system only through the
invariant mass $M$ (which in turn depends on the relative
variables and on the type of interaction).

\bigskip

B) The internal description concerns the relative degrees of
freedom of the isolated system inside the Wigner hyper-plane
(replacing the absolute Newtonian Euclidean 3-space containing the
isolated system). In order  to avoid a double counting of the
center-of-mass degrees of freedom there is the \textit{rest-frame
condition}, which implies the existence of the following 3 first
class constraints on the internal 3-momentum ${\vec{p}} =
\sum_{i=1}^{N}\, {\vec{\kappa}}_{i} \approx 0$ \footnote{ In the
non-relativistic limit we get the standard description with
${\vec{ \eta}}_{i} = {\vec{x}}_{i}$, ${\vec{\kappa}}_{i} =
{\vec{p}}_{i}$ in the Newtonian rest frame $\vec{p} \approx 0$.}.
This implies that a collective 3-variable (the \textit{internal}
3-center of mass) inside each Wigner hyper-plane can be
eliminated, so that only $3N-3$ internal relative canonical
variables are independent.\medskip

Since the sin tensor ${\bar S}^{AB}$ satisfies a Lorentz algebra
(see footnote 7), we can build an \textit{unfaithful internal}
realization of the Poincar\'{e} algebra, acting inside the Wigner
hyperplane, by adding the internal 3-momentum $p^r =
\epsilon^r_{\mu}(u(P))\, P^{\mu}_{sys} \approx 0$ and the
invariant mass $M = \sqrt{P^2_{sys}}$ as the internal energy. The
internal Poincare' generators $p^{\tau} = M$, $p^r$, $j^r = {\bar
S}^r$, $k^r = {\bar S}^{\tau r}$ can also be found from the
energy-momentum tensor of the isolated system evaluated in the
associated parametrized Minkowski theory and then restricted to
the rest-frame instant form.

The internal Poincare' generators for N free particles are
\cite{2}

\begin{eqnarray}
&&M = p^{\tau} = \sum_{i=1}^N\, \sqrt{m^2_i + {\vec \kappa}_i^2},  \notag \\
&&{\vec p} = \sum_{i=1}^N\, {\vec \kappa}_i\, (\approx 0),  \notag \\
&&\vec j = \sum_{i=1}^N\, {\vec \eta}_i \times {\vec \kappa}_i,\quad\quad
j^r = {\bar S}^r = {\frac{1}{2}}\, \epsilon^{ruv}\, {\bar S}^{uv},  \notag \\
&&\vec k = - \sum_{i=1}^N\, \sqrt{m^2_i + {\vec \kappa}_i^2}\,\,
{\vec \eta} _i,\quad\quad k^r = j^{or} = {\bar S}^{o r}.
  \label{7}
\end{eqnarray}

They satisfy the Poincare' algebra (like the external ones)

\begin{eqnarray}
\{ p^{\tau }, p_i \} &=& \{ p_i, p_j \} = 0,  \notag \\
\{ p_{i}, k_{j} \} &=& \delta _{ij}\, p^{\tau } = \delta_{ij}\, M,  \notag \\
\{ p^{\tau }, k_j \} &=& \{p^{\tau }, j_i \}= 0, \notag \\
\{ j_{i}, j_{j} \} &=& \varepsilon _{ijk}\, j_{k},  \notag \\
\{ j_{i}, k_{j}\} &=& \varepsilon _{ijk}\, k_{k},  \notag \\
\{ k_{i}, k_{j} \} &=& - \varepsilon _{ijk}\, j_{k}.
  \label{8}
\end{eqnarray}

\noindent The Poisson brackets $\{ p_i, k_j \} = \delta_{ij}\,
p^{\tau}$ shows clearly that the presence of interaction
potentials in the invariant mass $p^{\tau} = M$ requires the
presence of potentials also in the boost generators $k_i$.

\bigskip

Let us remark that, since we have ${\bar S}^{AB} =
\epsilon^A_{\mu}(u(P))\, \epsilon^B_{\nu}(u(P))\, \Big(J^{\mu\nu}
- x^{\mu}_s\, P^{\nu} + {\tilde x}^{\nu}\, P^{\mu}\Big)$, then
Eq.(\ref{4}) implies ${\bar S}^{rs} = \epsilon^r_{\mu}(u(P))\,
\epsilon^s_{\nu}(u(P))\, J^{\mu\nu}$ and ${\bar S}^{\tau r} =
u_{\mu}(P)\, \epsilon^r_{\nu}(u(P))\, J^{\mu\nu} + M\,
\epsilon^r_{\nu}(u(P))\, x^{\nu}_s(0)$. Therefore we have ${\bar
S}^{\tau r} = u_{\mu}(P)\, \epsilon^r_{\nu}(u(P))\, J^{\mu\nu}$
only with the choice $x^{\mu}_s(0) = 0$, i.e. when $x^{\mu}_s(\tau
) = u^{\mu}(P)\, \tau$. Moreover, as said in footnote 3, $k^r =
{\bar S}^{\tau r}$ is a gauge variables, whose natural gauge
fixing is $\vec k \approx 0$.

\bigskip

In Ref.\cite{27} we found the generators for a system of
$N-$interacting Grassmann charged particles and electromagnetic
fields. We started from the standard classical action reexpressed
as a parametrized Minkowski theory. Then with a suitable canonical
transformation we decoupled the scalar and longitudinal gauge
degrees of freedom of the electromagnetic field: in this way we
obtained a canonical formulation of the radiation gauge (only the
transverse vector potentials and electric fields are present) with
the emergence of the action-at-a-distance Coulomb potentials among
the charged particles.\ For $N = 2$ the internal Hamiltonian and
boosts have the following form   [$c(\vec \sigma ) = - 1/4\pi\,
|\vec \sigma |$]

\begin{eqnarray}
M &=& \sqrt{m_1^2 + ({\vec \kappa}_1(\tau ) - Q_1\, {\vec
A}_{\perp }(\tau , \vec{\eta}_1(\tau )))^2} + \sqrt{ m_2^2 +
({\vec \kappa }_2(\tau ) - Q_2\, {
\vec A}_{\perp }(\tau ,\vec{ \eta}_2(\tau )))^2} +  \notag \\
&+& \frac{Q_{1}\, Q_{2}}{4\pi\, \mid \vec{\eta}_{1}(\tau
)-\vec{\eta} _{2}(\tau )\mid } + \int d^3\sigma\, {\frac{1}{2}}\,
[{\vec E }_{\perp }^2 + {\vec B} ^2](\tau ,\vec{\sigma}),
 \nonumber \\
 &&{}\nonumber \\
 k^r &=& - \sum_{i=1}^2\, \eta^r_i(\tau )\, \sqrt{m^2_i +
({\vec \kappa}_i(\tau ) - Q_i\, {\vec A}_{\perp }(\tau ,
\vec{\eta}_i(\tau )))^2} +\nonumber \\
 &+& \sum_{i=1}^2\, \Big[ Q_1\, Q_2\, \sum_{i\not= j}\,
\Big({1\over {\nabla_{{\vec \eta}_j}}}\,\, {{\partial}\over
{\partial \eta^r_j}}\, c({\vec \eta}_i(\tau ) - {\vec \eta}_j(\tau ))
- \eta^r_j(\tau )\, c({\vec \eta}_i(\tau ) - {\vec \eta}_j(\tau ))
\Big) +\nonumber \\
 &+& Q_i\, \int d^3\sigma\, E^r_{\perp}(\tau ,\vec \sigma )\,
c(\vec \sigma - {\vec \eta}_i(\tau ))\Big] - {1\over 2}\, \int
d^3\sigma\, \sigma^r\, [{\vec E}^2_{\perp} + {\vec B}^2](\tau
,\vec \sigma ),
  \label{9}
\end{eqnarray}

\medskip

In the sector without an independent radiation field it can be
shown \cite{27} that the Coulomb potential is replaced by the
classical Darwin potential. The final form of the internal
generators $M$ and $\vec k$ in this sector is given in Eqs.(6.19)
and (6.46) of Ref.\cite{27}, respectively. Let us remark that
starting from classical electrodynamics we arrive at a Coulomb
potential additive to the square roots, and not living inside them
like in the toy model at the end of Appendix A, whose rest-frame
instant form will be studied in Section VI.

\vfill\eject

\section{The Problem of the Relativistic Center of Mass.}

As shown in Refs.\cite{2}, given an isolated system with an associated
realization of the Poincare' algebra, only three notions of collective
3-variables (coinciding only in the rest frame) can be built in term of them
(namely without introducing external variables). This is done by using the
group theoretical methods of Refs.\cite{5}. Then, these 3-variables have to
be extended to suitable collective 4-variables. They are

\noindent  i) a non-canonical non-covariant M\o ller
\textit{center of energy} \cite{8}, defining a frame-dependent
pseudo-world-line (it is the non-relativistic prescription with
the particle energies replacing their masses);\hfill\break
 ii) a canonical non-covariant center of mass (or
\textit{center of spin}); it is the classical analogue
\cite{10,11} of the Newton-Wigner position operator \cite{7} and
defines a frame-dependent pseudo-world-line; \hfill\break
 iii) a non-canonical covariant Fokker-Pryce \textit{center of inertia}
\cite {9,10}, leading to a 4-vector defining a frame-independent
world-line. \hfill\break

However, no-one of these candidates to represent the relativistic center of
mass has all the properties of the non-relativistic center of mass. \medskip

Since in the rest-frame instant form of dynamics we have both an
\textit{ internal} and an \textit{external} realization of the
Poincare' algebra, in this Section we shall review the definition
of the three collective variables in both cases starting from the
internal one.\bigskip

The internal realization of the Poincare' algebra leads to the following
internal (Wigner spin-1) collective 3-variables inside the Wigner
hyper-planes (therefore they need not to be extended to 4-variables):

\noindent i) a non-canonical \textit{internal} M\o ller
\textit{center of energy} ${\vec R}_{+}$;\hfill\break
 ii) a canonical \textit{internal} center of mass (or
\textit{center of spin}) ${\vec q}_{+}$; \hfill\break
  iii) a non-canonical \textit{internal} Fokker-Pryce \textit{center of
inertia} ${\vec y}_{+}$. \hfill\break

We shall see that on the Wigner hyper-planes, due to the rest
frame condition ${\vec p}\approx 0$, all of them coincide: ${\vec
q}_{+} \approx { \vec R}_{+} \approx {\vec y}_{+}$. \hfill\break

The 3-variables ${\vec R}_{+}$, ${\vec q}_{+}$, ${\vec y}_{+}$ have the
following definitions: \medskip

i) The \textit{internal} Moller 3-center of energy and the associated spin
vector are

\begin{eqnarray}
{\vec R}_{+}&=& - {\frac{1}{{M}}} \vec k = {\frac{{\sum_{i=1}^N\,
\sqrt{ m^2_i + {\vec \kappa}^2_i}\,\, {\vec
\eta}_i}}{{\sum_{k=1}^N\, \sqrt{m_k^2 + {
\vec \kappa}_k^2}}}},  \notag \\
{\vec S}_R &=& \vec j -{\vec R}_{+} \times {\vec p},  \notag \\
&&{}  \notag \\
&&\{ R^r_{+}, p^s \} = \delta^{rs},\qquad \{ R^r_{+}, M \} =
{\frac{{p^r}}{{M }}},\qquad \{ R^r_{+}, R^s_{+} \} = -
{\frac{1}{{M^2}}}\, \epsilon^{rsu}\,
S_R^u,  \notag \\
&&\{ S_R^r, S_R^s \} = \epsilon^{rsu}\, (S_R^u - {\frac{1}{{M^2}}}
{\vec S_R} \cdot {\vec p}\,\, p^u),\quad\quad \{ S_R^r ,M \} =0.
  \label{10}
\end{eqnarray}

Therefore, the \textit{internal} boost generator may be rewritten as $\vec k
= - M\, {\vec R}_{+}$, so that ${\vec R}_{+} \approx 0$ implies $\vec k
\approx 0$ (see footnote 3).

Note that in the non-relativistic limit ${\vec R}_{+}$ tends the
the non-relativistic center of mass ${\vec q}_{nr}=
{\frac{{\sum_{i=1}^N\, m_i\, {\vec \eta}_i}}{{\sum_{i=1}^N\,
m_i}}}$.

\medskip

ii) The canonical \textit{internal} 3-center of mass and the associated spin
vector are

\begin{eqnarray}
{\vec q}_{+}&=& - {\frac{{\vec k}}{\sqrt{M^2 - {\vec p}^2}}} +
{\frac{{\ \vec j \times {\vec p}}}{{\sqrt{M^2 - {\vec p}^2}\, (M +
\sqrt{M^2 - {\vec p}^2})}}} +  \notag \\
&+&{\frac{{\vec k \cdot {\vec p}\,\, {\vec p}}}{{M\, \sqrt{M^2 -
{\vec p}^2}
\, \Big( M + \sqrt{M^2 - {\vec p}^2}\Big) }}},  \notag \\
&&{}  \notag \\
&&\approx {\vec R}_{+}\quad for\quad {\vec p} \approx 0;\qquad \{
{\vec q}
_{+}, M \} = {\frac{{\vec p}}{{M}}},  \notag \\
&&{}  \notag \\
{\vec S}_q &=&\vec j - {\vec q}_{+} \times {\vec p} = {\frac{{M\,
\vec j}}{
\sqrt{M^2 - {\vec p}^2}}}+  \notag \\
&+&{\frac{{\ \vec k \times {\vec p}}}{\sqrt{M^2 - {\vec p}^2}}} -
{\frac{{ \vec j \cdot {\vec p}\,\, {\vec p}}}{{\sqrt{M^2 - {\vec
p}^2}\, \Big( M +
\sqrt{M^2 - {\vec p}^2}\Big) }}} \approx {\vec {\bar S}} = \vec j,  \notag \\
&&{}  \notag \\
&&\{ {\vec S}_q, {\vec p} \} = \{ {\vec S}_q, {\vec q}_{+} \} = 0, \qquad \{
S_q^r, S_q^s \} = \epsilon^{rsu}\, S_q^u.
  \label{11}
\end{eqnarray}

\medskip

iii) The \textit{internal} non-canonical Fokker-Pryce center of
inertia ${ \vec y}_{+}$ is

\begin{eqnarray}
{\vec y}_{+}&=& {\vec q}_{+} + {\frac{{\ {\vec S}_q \times {\vec
p}}}{{\sqrt{ M^2 - {\vec p}^2}\, (M + \sqrt{M^2 - {\vec p}^2})}}}
= {\vec R}_{+} + {\frac{
{\ {\vec S}_q \times {\vec p}}}{{M\, \sqrt{M^2 - {\vec p}^2}}}},  \notag \\
&&{}  \notag \\
&&\{ y^r_{+}, y^s_{+} \} = {\frac{1}{{M\, \sqrt{M^2 - {\vec p}^2}
}}}\, \epsilon^{rsu}\,\Big[ S^u_q + {\frac{{\ {\vec S}_q \cdot
{\vec p}\, p^u} }{{
\sqrt{M^2 - {\vec p}^2}\, (M + \sqrt{M^2 - {\vec p}^2})}}}\Big].  \notag \\
&&{}
 \label{12}
\end{eqnarray}

\medskip

We have

\begin{eqnarray}
{\vec q}_{+}&=&{\vec R}_{+} + {\frac{{\ {\vec S}_q \times {\vec p}}}{{M\, (M
+ \sqrt{M^2 - {\vec p}^2})}}} = {\frac{{M\, {\vec R}_{+} + \sqrt{M^2 - {\vec
p}^2}\, {\vec y}_{+}}}{{M + \sqrt{M^2 - {\vec p}^2}}}},  \notag \\
&&{}  \notag \\
{\vec p} \approx 0 &\Rightarrow& {\vec q}_{+} \approx {\vec R}_{+}
\approx { \vec y}_{+}.
  \label{13}
\end{eqnarray}

We see that the gauge fixings ${\vec q}_{+} \approx {\vec R}_{+}
\approx {\vec y} _{+} \approx 0$ force the three \textit{internal}
collective 3-variables to coincide with the location of the
inertial observer $x^{\mu}_s(\tau )$, origin of the 3-coordinates
(see Refs.\cite{2} for other properties of $ x^{\mu}_s(\tau )$ in
this gauge).

\bigskip

On the other hand from the external realization of the Poincare'
algebra we get the following three \textit{external} collective
3-variables (the canonical ${ \vec q}_s$, the Moller ${\vec R}_s$
and the Fokker-Pryce ${\vec Y} _s$)

\begin{eqnarray}
{\vec R}_s &=& - {\frac{{1}}{{P^o}}}\, {\vec K} = ({\vec {\tilde
x}} - {\frac{ {\vec P}}{{P^o}}}\, {\tilde x}^o) - {\frac{{{\vec
{\bar S}} \times {\vec P}}
}{{P^o\, (P^o + M)}}},  \notag \\
{\vec q}_s &=& {\vec {\tilde x}} - {\frac{{\vec P}}{{P^o}}}\,
{\tilde x}^o = { \vec R}_s + {\frac{{\ {\vec {\bar S}} \times
{\vec P}}}{{P^o\, (P^o + M)}}}
= {\frac{{P^o\, {\vec R}_s + M\, {\vec Y}_s}}{{P^o + M}}} ,  \notag \\
{\vec Y}_s&=&{\vec q}_s + {\frac{{\ {\vec {\bar S}} \times {\vec
P}}}{{M\, (P^o + M)}}} = {\vec R}_s + {\frac{{\ {\vec {\bar S}}
\times {\vec P}}}{{P^o\, M}}},  \notag \\
&&{}  \notag \\
&&\{ R^r_s, R^s_s \} = - {\frac{{1}}{{(P^o)^2}}}\,
\epsilon^{rsu}\, \Omega^u_s, \quad\quad {\vec \Omega}_s = {\vec J}
- {\vec R}_s \times {\vec P
},  \notag \\
&&{}  \notag \\
&&\{ q^r_s, q^s_s \} =0,\qquad \{ Y^r_s, Y^s_s \} = {\frac{1}{{M\,
P^o}}}\, \epsilon^{rsu}\, \Big[ {\bar S}^u + {\frac{{\ {\vec {\bar
S}} \cdot {\vec P}
\, P^u}}{{M\, (P^o + M)}}}\Big] ,  \notag \\
&&{}  \notag \\
{\vec P} \cdot {\vec q}_s &=& {\vec P} \cdot {\vec R}_s = {\vec P}
\cdot {\vec Y}_s,  \notag \\
&&{}  \notag \\
{\vec P} = 0 &\Rightarrow& {\vec q}_s = {\vec Y}_s = {\vec R}_s.
  \label{14}
\end{eqnarray}

\noindent All of them have the same velocity and coincide in the
Lorentz rest frame where ${\buildrel \circ \over P}^{\mu} = M\,
(1;\vec 0)$
\medskip

As shown in Refs.\cite{2}, the requirement that the relations
$\tau \equiv T_{s} = u(P) \cdot x_{s} = u(P) \cdot {\tilde{x}} =
u(P) \cdot Y_{s} = u(P) \cdot R_{s}$ holds on Wigner hyper-planes
allows us to extend the external collective 3-variables to the
following external 4-variables: \medskip

\noindent i) the \textit{external} non-canonical and non-covariant
M\o ller 4-center of energy $R^{\mu }_s$ (a frame-dependent
pseudo-world-line, whose intersection with the Wigner hyper-plane
has 3-coordinate $\sigma_R^r$); \hfill\break
 ii) the \textit{external} canonical non-covariant
4-center of mass ${\tilde x} ^{\mu }$ (a frame-dependent
pseudo-world-line, whose intersection with the Wigner hyper-plane
has 3-coordinate ${\tilde \sigma}^r$); \hfill\break
 iii) the \textit{external} covariant non-canonical Fokker-Pryce 4-center of
inertia $Y^{\mu }_s$ (a frame-independent world-line, whose
intersection with the Wigner hyper-plane has 3-coordinate
$\sigma_Y^r$).

They have the following definitions \footnote{ As shown in
Refs.\cite{2,21} the canonical variables ${\tilde{x}}^{\mu }$, $
P^{\mu }$, can be replaced by the new canonical basis $T_{s} =
u(P) \cdot \tilde{x}$, $\epsilon _{s} = \sqrt{\,P^{2}}$, $z^{i} =
\sqrt{\,P^{2}}\, ({\tilde{x }}^{i} - {\frac{{P^{i}}}{{P^{o}}}}\,
{\tilde{x}}^{o})$, $h^{i} = P^{i}/\sqrt{ \,P^{2}}$ restricted by
the first class constraint $\epsilon _{s} - M \approx 0$ . The
gauge fixing $T_{s} - \tau \approx 0$ (identifying $\tau $ with
the Lorentz-scalar rest time) implies $T_{s} \equiv \tau $,
$\epsilon _{s} \equiv M $, at the level of Dirac brackets. The
3-vector $\vec{z}/M$ is the non-covariant canonical 3-center of
mass (the classical counterpart of the ordinary Newton-Wigner
position operator), which has 3-velocity $\vec{h}$ (or 3-momentum
$M\, \vec{h}$). Note that $\vec{z}$ and $\vec{h}$ are non-evolving
Jacobi data. In the text we used ${\tilde x}^{\mu}$ and $P^{\mu}$
instead of their expressions in terms of $\vec z$ and $\vec h$
(${\tilde x}^o = \sqrt{1 + {\vec h}^2}\, (\tau + {{\vec h \cdot
\vec z}\over {M}})$, ${\vec {\tilde x}} = {{\vec z}\over M} +
(\tau + {{\vec h \cdot \vec z}\over {M}})\, \vec h$, $P^o = M\,
\sqrt{1 + {\vec h}^2}$, $\vec P = M\, \vec h$) to simplify the
notation. The external Poincare' algebra (\ref{6}) is satisfied
with the following form of the generators: $P^o = M\, \sqrt{1 +
{\vec h}^2}$, $\vec P = M\, \vec h$, $J^{ij} = z^i\, h^j - z^j\,
h^i + {\bar S}^{ij}$, $K^i = - \sqrt{1 + {\vec h}^2}\, z^i +
{{\epsilon^{irs}\, h^r\, {\bar S}^s}\over {1 + \sqrt{1 + {\vec
h}^2}}}$.}

\begin{eqnarray}
{\tilde x}^{\mu} &=&( {\tilde x}^o; {\vec {\tilde x}}) = ({\tilde
x}^o; { \vec q}_s + {\frac{{\vec p}}{{\ p^o}}}\, {\tilde x}^o)
\equiv x^{\mu}_s +
\epsilon^{\mu}_u(u(P))\, {\tilde \sigma}^u,  \notag \\
&&{}  \notag \\
Y^{\mu}_s&=& ({\tilde x}^o; {\vec Y}_s) =  \notag \\
&=&{\tilde x}^{\mu} + \epsilon^{\mu}_r(P)\, {\frac{{({\vec {\bar
S}} \times { \vec P})^r}}{{M\, [1 + u^o(P)]}}} \equiv x^{\mu}_s +
\epsilon^{\mu}_u(u(P))\, \sigma^u_Y,  \notag \\
&&{}  \notag \\
R^{\mu}_s&=&({\tilde x}^o; {\vec R}_s) =  \notag \\
&=&{\tilde x}^{\mu} - \epsilon^{\mu}_r(P)\, {\frac{{({\vec {\bar
S}} \times { \vec P})^r} }{{M\, u^o(P)\, [1 + u^o(P)]}}} \equiv
x^{\mu}_s + \epsilon^{\mu}_u(u(P))\, \sigma^u_R,
  \label{15}
\end{eqnarray}

\bigskip

\noindent and we have

\begin{eqnarray}
\sigma^r_Y&=& \epsilon_{r\mu}(u(P))\, [x^{\mu}_s - Y^{\mu}_s] = {\tilde
\sigma}^r + {\frac{{{\bar S}^{rs}\, u^s(P)}}{{1 + u^o(P)}}} =  \notag \\
&=& M\, R^r_{+} \approx M\, q^r_{+} \approx 0,  \notag \\
&&{}  \notag \\
&\Rightarrow& x^{\mu}_s(\tau ) = x^{\mu}(0) + Y^{\mu}_s(\tau ),\,
\mathit{ when}\,\,\, {\vec q}_{+}\approx 0,
  \label{16}
\end{eqnarray}

\noindent As a consequence the gauge fixings ${\vec q}_{+} \approx
{\vec R} _+ \approx {\vec y}_+ \approx 0$ implies that we can
choose the inertial observer $x^{\mu }_s$ to coincide with the
covariant (non-canonical) \textit{ external Fokker-Pryce center of
inertia} by choosing $x^{\mu}(0) = 0$.
\medskip

Let us remark that in the inertial rest-frame instant form with
the gauge fixings $\vec k \approx 0$, i.e. ${\vec q}_{+} \approx
{\vec R}_+ \approx { \vec y}_+ \approx 0$, the only non-zero
generators of the internal Poincare' algebra are $M$ and $\vec j =
{\vec {\bar S}}$: they contain all the information about the
isolated system and generate the dynamical U(2) algebra of
Ref.\cite{29}. As is evident from Eqs.(\ref{6}) also the external
Poincare' generators depend only on the generators of this U(2)
algebra.

Since we are in an instant form of dynamics, in the presence of
interactions among the constituents of the isolated system only
the internal generators $ M $ and $\vec{k}$ will contain the
interaction potentials, $M \mapsto M_{(int)}$ , $\vec{k} \mapsto
{\vec{k}}_{(int)}$, but only the ones inside $M_{(int)}$
contribute to the U(2) algebra. As shown in the next Section, the
potentials inside ${\vec{k}}_{(int)}$ contribute to the
elimination of the internal 3-centers by means of the gauge
fixings ${\vec{k}}_{(int)} \approx 0$.

\medskip As anticipated in Section II, in each Lorentz frame one has
different pseudo-world-lines describing $R_{s}^{\mu }$ and
${\tilde{x}} _{s}^{\mu }$: the canonical 4-center of mass
${\tilde{x}}_{s}^{\mu }$ lies nearer to $Y_{s}^{\mu }$ than
$R_{s}^{\mu }$ in every (non rest-) frame. In an arbitrary Lorentz
frame, the pseudo-world-lines associated with ${\tilde{x
}}_{s}^{\mu }$ and $R_{s}^{\mu }$ fill a M$\o $ller world-tube
around the world-line $Y_{s}^{\mu }$ of the covariant
non-canonical Fokker-Pryce 4-center of inertia $Y_{s}^{\mu }$.

\vfill\eject

\section{The Canonical Transformation to Relative Variables.}

From the previous Sections it is clear that in the rest-frame
instant form the N-body problem in the free case is described by
the $8 + 6N$ canonical variables ${\tilde x}$, $P^{\mu}$, ${\vec
\eta}_i$, ${\vec \kappa}_i$, $i=1,..,N$, restricted by the 6
conditions $\vec p \approx 0$, ${\vec q}_+ \approx 0$ and with
$P^{\mu} = M\, u^{\mu}(P)$, $M = \sum_{i=1}^N\, \sqrt{m^2_i +
{\vec \kappa}^2_i}$, and $\tau \equiv T_s = u(P) \cdot \tilde x$,
so that there are only $6 + 6\, (N-1)$ independent canonical
variables [$\vec z$, $\vec h$, ${\vec \rho}_{qa}$, ${\vec
\pi}_{qa}$, $a=1,..,N-1$] like in the non-relativistic case (see
footnote 9 for the definition of $\vec z$ and $\vec h$).

We have now to find the canonical transformation

\begin{eqnarray}
&&\begin{minipage}[t]{1cm} \begin{tabular}{|l|l|} \hline ${\tilde x}^{\mu}$
& ${\vec \eta}_i$ \\ \hline $P^{\mu}$ & ${\vec \kappa}_i$ \\ \hline
\end{tabular} \end{minipage} \ {\longrightarrow \hspace{.2cm}} \ %
\begin{minipage}[t]{2 cm} \begin{tabular}{|l|l|l|} \hline ${\tilde x}^{\mu}$
& ${\vec q}_{+}$ & ${\vec \rho}_{qa}$ \\ \hline $P^{\mu}$ & ${\vec
p}$&${\vec \pi}_{qa}$ \\ \hline \end{tabular} \end{minipage},  \notag \\
&&{}  \notag \\
&& \vec p \approx 0,\qquad {\vec q}_+ \approx 0,
  \label{17}
\end{eqnarray}

\noindent defining the $6\, (N-1)$ relativistic relative variables
${\vec \rho}_{q,a}$, ${\vec \pi}_{q,a}$, $a=1,..,N-1$, so that the
spin (barycentric angular momentum) becomes ${\vec S}_q =
\sum_{a=1}^{N-1}\, {\vec \rho}_{q,a} \times {\vec \pi}_{q,a}$.
\medskip

Let us stress that this cannot be a point transformation, because
of the momentum dependence of the relativistic internal center of
mass ${\vec q} _{+} $. \bigskip

Since ${\vec{q}}_{+}$ and ${\vec{p}}$ are known from Eqs.
(\ref{11}) and ( \ref{7}) respectively, we have only to find the
internal conjugate variables appearing in the canonical
transformation (\ref{17}). They have been determined in
Ref.\cite{2} by using the technique (the Gartenhaus-Schwarz
transformation) of Ref.\cite{30} and starting from a set of
canonical variables defined in Ref.\cite{21}. Then, starting from
the naive \textit{internal} center-of-mass variable
${\vec{\eta}}_{+} = {\frac{1}{N}}\, \sum_{i=1}^{N}\,
{\vec{\eta}}_{i}$, we defined relative variables
${\vec{\rho}}_{a}$, ${\vec{\pi}}_{a}$ based on the following
family of point canonical transformations

\begin{eqnarray*}
&&\begin{minipage}[t]{3cm} \begin{tabular}{|l|} \hline ${\vec \eta}_i$ \\
\hline ${\vec \kappa}_i$ \\ \hline \end{tabular} \end{minipage} \
{ \longrightarrow \hspace{.2cm}} \ \begin{minipage}[t]{2 cm}
\begin{tabular}{|l|l|} \hline ${\vec \eta}_{+}$ & ${\vec \rho}_a$ \\ \hline
${\vec p}$&${\vec \pi}_a$ \\ \hline \end{tabular} \end{minipage} ,
\quad\quad a=1,..,N-1,
\end{eqnarray*}

\begin{eqnarray}
 {\vec{\eta}}_{+} &=&{\frac{1}{N}}\,\sum_{i=1}^{N}\, {\vec{\eta}}_{i},\qquad {
\vec{p}} = \sum_{i=1}^{N}\, {\vec{\kappa}}_{i} \approx 0,  \notag \\
{\vec{\rho}}_{a} &=& \sqrt{N}\, \sum_{i=1}^{N}\, \gamma _{ai}\,
{\vec{\eta}} _{i},\qquad {\vec{\pi}}_{a} = {\frac{1}{\sqrt{N}}}\,
\sum_{i=1}^{N}\, \gamma_{ai}\, {\vec{\kappa}}_{i},  \notag \\
&&{}  \notag \\
&&\{\eta _{i}^{r}, \kappa _{j}^{s}\} = \delta _{ij}\, \delta
^{rs},\quad \quad \{\eta _{+}^{r}, p^{s}\} = \delta ^{rs},\quad
\quad \{\rho _{a}^{r}, \pi _{b}^{s}\} = \delta _{ab}\, \delta
^{rs},\nonumber \\
 &&{}\nonumber \\
{\vec{\eta}}_{i} &=& {\vec{\eta}}_{+} + {\frac{1}{\sqrt{N}}}\,
\sum_{a=1}^{N-1} \,\gamma _{ai}\, {\vec{\rho}}_{a},\qquad
{\vec{\kappa}}_{i} = {\frac{1}{N}}\, { \vec{p}} + \sqrt{N}\,
\sum_{a=1}^{N-1}\, \gamma_{ai}\, {\vec{\pi}}_{a},\nonumber \\
 &&{}
  \label{18}
\end{eqnarray}
In order that the above brackets be satisfied, the numerical
parameters $ \gamma _{ai}$ must satisfy the relations
$\sum_{i=1}^{N}\, \gamma _{ai} = 0$, $ \sum_{i=1}^{N}\, \gamma
_{ai}\, \gamma _{bi} = \delta _{ab}$, $ \sum_{a=1}^{N-1}\, \gamma
_{ai}\, \gamma _{aj} = \delta _{ij} - {\frac{1}{N}}$, which depend
on ${\frac{1}{2}}(N-1)(N-2)$ free parameters.

Then, (in Appendix B of Ref.\cite{31}), we gave the closed form of
the canonical transformation (\ref{17}) for arbitrary N, which
turned out to be \textit{ point in the momenta} but, unlike the
non-relativistic case, \textit{ non-point} in the configurational
variables. \medskip

Explicitly, for $N=2$ we have [$\gamma_{11} = - \gamma_{12} =
{\frac{1}{ \sqrt{2}}}$, so that $\vec \rho = {\vec \eta}_1 - {\vec
\eta}_2$, $\vec \pi = { \frac{1}{2}}\, ({\vec \kappa}_1 - {\vec
\kappa}_2)$]

\begin{eqnarray}
M &=& \sqrt{m_1^2 + {\vec \kappa}_1^2} + \sqrt{m_1^2 + {\vec
\kappa}_1^2}
,\qquad {\vec S}_q = {\vec \rho}_q \times {\vec \pi}_q,  \notag \\
&&{}  \notag \\
{\vec q}_+ &=& {\frac{{\sqrt{m_1^2 + {\vec \kappa}_1^2}\, {\vec
\eta}_1 + \sqrt{m_2^2 + {\vec \kappa}_2^2}\, {\vec
\eta}_2}}{\sqrt{M^2 - {\vec p}^2}}} + {\frac{{({\vec \eta}_1
\times {\vec \kappa}_1 + {\vec \eta}_2 \times {\vec \kappa}_2)
\times {\vec p}}}{{\sqrt{M^2 - {\vec p}^2}\, (M + \sqrt{M^2 - {
\vec p}^2})}}} -  \notag \\
&-& {\frac{{(\sqrt{m_1^2 + {\vec \kappa}_1^2}\, {\vec \eta}_1 + \sqrt{m_2^2
+ {\vec \kappa}_2^2}\, {\vec \eta}_2) \cdot {\vec p}\,\, {\vec p}}}{{M\,
\sqrt{M^2 - {\vec p}^2}\, (M + \sqrt{M^2 - {\vec p}^2})}}},  \notag \\
{\vec p} &=& {\vec \kappa}_1 + {\vec \kappa}_2 \approx 0,  \notag \\
{\vec \pi}_q &=& \vec \pi - {\frac{{\vec p}}{\sqrt{M^2 - {\vec
p}^2}}}\, \Big[{\frac{1}{2}}\, (\sqrt{m_1^2 + {\vec \kappa}_1^2} -
\sqrt{m_2^2 + {\vec
\kappa}_2^2}) -  \notag \\
&-& {\frac{{{\vec p} \cdot \vec \pi}}{{{\vec p}^2}}}\,\, (M -
\sqrt{M^2 - { \vec p}^2})\Big] \approx \,\, \vec \pi =
{\frac{1}{2}}\, ({\vec \kappa}_1 - {
\vec \kappa}_2),  \notag \\
{\vec \rho}_q &=& \vec \rho + \Big({\frac{\sqrt{m_1^2 + {\vec
\kappa}_1^2}}{ \sqrt{m_2^2 + {\vec \pi}_q^2}}} +
{\frac{\sqrt{m_2^2 + {\vec \kappa}_2^2}}{ \sqrt{m_1^2 + {\vec
\pi}_q^2}}}\Big)\, {\frac{{{\vec p} \cdot \vec \rho\,\, { \vec
\pi}_q}}{{M\, \sqrt{M^2 - {\vec p}^2}}}} \approx \,\, \vec \rho =
{\vec \eta}_1 - {\vec \eta}_2,  \notag \\
&&{}  \notag \\
&&{}  \notag \\
\Rightarrow&& M = \sqrt{\mathcal{M}^2 + {\vec p}^2} \approx\,\, \mathcal{M}
= \sqrt{m_1^2 + {\vec \pi}_q^2} + \sqrt{m_2^2 + {\vec \pi}_q^2},  \notag \\
&&\qquad {\vec q}_+ \approx {\vec R}_+ \approx {\frac{{{\vec
\eta}_1\, \sqrt{ m^2_1 + {\vec \pi}^2} + {\vec \eta}_2\,
\sqrt{m^2_2 + {\vec \pi}^2}}}{{ \mathcal{M}}}}.
  \label{19}
\end{eqnarray}

${\cal M}$ is the rest-frame internal energy expressed in terms of
relative variables.

\medskip

The inverse canonical transformation is

\begin{eqnarray*}
{\vec \eta}_i &=& {\vec q}_+ - {\frac{{{\vec S}_q \times {\vec
p}}}{{\sqrt{ \mathcal{M}^2 + {\vec p}^2}\, (\mathcal{M} +
\sqrt{\mathcal{M}^2 + {\vec p}^2
}}}} + {\frac{1}{2}}\, \Big[ (-1)^{i+1} -  \notag \\
&-& {\frac{{2\, \mathcal{M}\, {\vec \pi}_q \cdot {\vec p} + (m_1^2
- m_2^2)\, \sqrt{\mathcal{M}^2 + {\vec p}^2}}}{{\mathcal{M}^2\,
\sqrt{\mathcal{ M}^2 + {\vec p}^2}}}}\Big] \cdot
\end{eqnarray*}

\begin{eqnarray}
&&\Big[{\vec \rho}_q - {\frac{{{\vec \rho}_q \cdot {\vec p}\,\,
{\vec \pi}_q} }{{\mathcal{M}\, \sqrt{\mathcal{M}^2 + {\vec p}^2}\,
\Big({\frac{\sqrt{m_1^2 + {\vec \kappa}_1^2}}{\sqrt{m_2^2 + {\vec
\pi}_q^2}}} + {\frac{\sqrt{m_2^2 + {\vec \kappa}_2^2}}{\sqrt{m_1^2
+ {\vec \pi}_q^2}}}\Big)^{-1} + {\vec \pi}_q
\cdot {\vec p}}}}\Big] \approx  \notag \\
&\approx& {\vec q}_+ + {\frac{1}{2}}\, \Big[ (-1)^{i+1} -
{\frac{{m_1^2 - m_2^2}}{{\mathcal{M}^2}}}\Big]\, {\vec \rho}_q
\approx {\frac{1}{2}}\, \Big[ (-1)^{i+1} - {\frac{{m_1^2 -
m_2^2}}{{\mathcal{M}^2}}}\Big]\, {\vec \rho},  \notag \\
&&{}  \notag \\
{\vec \kappa}_i &=& \Big[{\frac{1}{2}} +
{\frac{{(-1)^{i+1}}}{{\mathcal{M}\, \sqrt{\mathcal{M}^2 + {\vec
p}^2}}}}\, \Big({\vec \pi}_q \cdot {\vec p}\, [1 -
{\frac{{\mathcal{M}}}{{{\vec p}^2}}}\, (\sqrt{\mathcal{M}^2 +
{\vec p}^2}- \mathcal{M})] +  \notag \\
&+& (m_1^2 - m_2^2)\, \sqrt{\mathcal{M}^2 + {\vec
p}^2}\Big)\Big]\, {\vec p} + (-1)^{i+1}\, {\vec \pi}_q \approx
\,\, (-1)^{i+1}\, {\vec \pi}_q \approx
(-)^{i+1}\, \vec \pi,  \notag \\
\Rightarrow && {\vec \kappa}_i^2 \approx {\vec \pi}^2.
  \label{20}
\end{eqnarray}

In Eqs.(\ref{20}) we used explicitly the gauge fixing ${\vec q}_+
 \approx 0$.

\medskip

As shown in Refs.\cite{32}, \cite{21} and their bibliography,
a-a-a-d interactions inside the Wigner hyperplane may be
introduced  either under  (scalar and vector potentials) or
outside (scalar potential like the Coulomb one) the square roots
appearing in the free Hamiltonian. Since a Lagrangian density in
presence of action-at-a-distance mutual interactions is not known
and since we are working in an instant form of dynamics, the
potentials in the constraints restricted to hyper-planes must be
introduced \textit{by hand } [see, however, Ref.\cite{27} for
their evaluation starting from the Lagrangian density for the
electro-magnetic interaction]. The only restriction is that the
Poisson brackets of the modified constraints must generate the
same algebra of the free ones.

\medskip

In the rest-frame instant form the most general Hamiltonian with
action-at-a-distance interactions is

\begin{equation}
M_{(int)} = \sum_{i=1}^{N}\, \sqrt{m_{i}^{2} + U_{i} +
[{\vec{\kappa}}_{i} - {\vec{V}} _{i}]^{2}} + V,
  \label{21}
\end{equation}

\noindent where $U = U({\vec{\kappa}}_{k}, {\vec{\eta}}_{h} -
{\vec{\eta}}_{j\neq k})$, ${\vec{V}}_{i} =
{\vec{V}}_{i}({\vec{\kappa}}_{j\not=i}, {\vec{\eta}}_{i} -
{\vec{\eta}}_{j\not=i})$, $V = V_{o}(|{\vec{\eta}}_{i} -
{\vec{\eta}} _{j}|) + V^{^{\prime}}({\vec{\kappa}}_{i},
{\vec{\eta}}_{i} - {\vec{\eta}}_{j})$ .
\medskip

If we use the canonical transformation (\ref{17}) defining the
relativistic canonical internal 3-center of mass (now it is
interaction-dependent, ${\vec q}_+^{(int)}$) and relative
variables on the Wigner hyperplane, with the rest-frame conditions
$\vec p \approx 0$, the rest frame Hamiltonian for the relative
motion becomes

\begin{equation}
M_{(int)} \approx \sum_{i=1}^N\, \sqrt{m_i^2 + {\tilde U}_i +
[\sqrt{N}\, \sum_{a=1}^{N-1}\, \gamma_{ai}\, {\vec \pi}_{qa} -
{\tilde {\vec V}}_i]^2} + \tilde V,
  \label{22}
\end{equation}

\noindent where

\begin{eqnarray}
{\tilde U}_i &=& U([\sqrt{N}\, \sum_{a=1}^{N-1}\, \gamma_{ak}\,
{\vec \pi}_{qa}, {\frac{ 1}{\sqrt{N}}}\, \sum_{a=1}^{N-1}\,
(\gamma_{ah} - \gamma_{ak})\, {\vec \rho}_{qa}),
\notag \\
{\tilde {\vec V}}_i &=& {\vec V}_i([\sqrt{N}\, \sum_{a=1}^{N-1}\,
\gamma_{aj\not= i}\, {\vec \pi}_{qa}, {\frac{1}{\sqrt{N}}}\,
\sum_{a=1}^{N-1}\,
(\gamma_{ai} - \gamma_{aj\not= i})\, {\vec \rho}_{qa}),  \notag \\
\tilde V &=& V_o(|{\frac{1}{\sqrt{N}}}\, \sum_{a=1}^{N-1}\,
(\gamma_{ai} - \gamma_{aj})\, {\vec \rho}_{qa}|) +
V^{^{\prime}}([\sqrt{N}\, \sum_{a=1}^{N-1}\, \gamma_{ai}\, {\vec
\pi}_{qa}, {\frac{1}{\sqrt{N}}}\, \sum_{a=1}^{N-1}\, (
\gamma_{ai} - \gamma_{aj})\, {\vec \rho}_{qa}).  \notag \\
&&{}  \label{23}
\end{eqnarray}

\medskip

In order to build a realization of the internal Poincare' group,
besides $ M_{ (int)}$ we need to know the potentials appearing in
the internal boosts ${ \vec k}_{(int)}$ (being an instant form,
${\vec p} \approx 0$ and $\vec j$ are the free ones).

\bigskip

Since the 3-centers ${\vec R}_{+}$ and ${\vec q}_{+}$ become
interaction dependent, the final canonical basis ${\vec q}_{+}$,
${\vec p}$, ${\vec \rho} _{qa}$, ${\vec \pi}_{qa}$ is {\it not
explicitly known in the interacting case}. For an isolated system,
however, we have $M = \sqrt{\mathcal{M}^2 + {\vec p}^2} \approx
\mathcal{M}$ with $\mathcal{M}$ independent of ${\vec q}_{+}$ ($\{
M, {\vec p} \} = 0$ in the internal Poincare' algebra). This
suggests that the same result should hold true even in the
interacting case. Indeed, by its definition, the
Gartenhaus-Schwartz transformation \cite{30}, \cite{2} gives
${\vec \rho}_{qa} \approx {\vec \rho}_a$, ${\vec \pi}_{qa} \approx
{ \vec \pi}_a$ also in presence of interactions, so that we get

\begin{eqnarray}
M_{ (int)}{|}_{\vec p =0} &=& \Big( \sum_i\, \sqrt{m_i^2 + U_i +
({ \vec \kappa}_i - {\vec V}_i)^2} + V \Big){|}_{{\vec p} =0} =
\sqrt{\mathcal{M}_{(int)}^2 + {\vec p}^2} {|}_{\vec p = 0} =  \notag \\
&=& \mathcal{M}_{(int)}{|}_{{\vec p} =0} = \sum_i\, \sqrt{m_i^2 +
{\tilde U}_i + ({\vec \kappa}_i - {\vec {\tilde V}}_i)^2} + \tilde
V,
  \label{24}
\end{eqnarray}

\noindent where the potentials ${\tilde U}_i$, ${\vec {\tilde
V}}_i$, $ \tilde V$ are now functions of ${\vec \pi}_{qa} \cdot
{\vec \pi}_{qb}$, ${ \vec \pi}_{qa} \cdot {\vec \rho}_{qb}$,
${\vec \rho}_{qa} \cdot {\vec \rho} _{qb}$.

\bigskip

Unlike in the non-relativistic case, the canonical transformation
(\ref{19}) is now \textit{interaction dependent} (not even a point
transformation in the momenta), since ${\vec q}_+$ is determined
by a set of Poincare' generators depending on the interactions.
The only thing to do in the generic situation is therefore to use
the free relative variables (\ref{19}) even in the interacting
case. We cannot impose anymore, however, the natural gauge fixings
${\vec q}_+ \approx 0$ ($\vec k \approx 0$) of the free case,
since it is replaced by ${\vec q}_+^{(int)} \approx 0$ (namely by
${\vec k}_{(int)} \approx 0$), the only gauge fixing identifying
the centroid with the external Fokker-Pryce 4-center of inertia
also in the interacting case. Once written in terms of the
canonical variables (\ref{19}) of the free case, the equations
${\vec k}_{(int)} \approx 0$ can be solved for ${\vec q}_+$, which
takes a form ${\vec q} _+ \approx \vec f({\vec \rho}_{aq}, {\vec
\pi}_{aq})$ as a consequence of the potentials appearing in the
boosts. Therefore, for $N=2$, the reconstruction of the
relativistic orbit by means of Eqs.(\ref{20}) in terms of the
relative motion is given by (similar equations hold for arbitrary
N; $ {\vec \rho}_{aq} \approx {\vec \rho}_a$, ${\vec \pi}_{aq}
\approx {\vec \pi}_a$)

\begin{eqnarray}
{\vec \eta}_i(\tau ) &\approx& {\vec q}_+({\vec \rho}_q, {\vec
\pi}_q) + { \frac{1}{2}}\, \Big[ (-)^{i+1} - {\frac{{m_1^2 -
m_2^2}}{{\mathcal{M}^2}}} \Big]\, {\vec \rho}_q {\rightarrow}_{c
\rightarrow \infty}\, {\frac{1}{2}}
\Big[(-)^{i+1} - {\frac{{m_1-m_2}}{{m}}}\Big]\, {\vec \rho}_q,  \notag \\
&&{}  \notag \\
{\vec \kappa}_i(\tau ) &\approx& (-)^{i+1}\, {\vec \pi}_{q}(\tau
),  \notag \\
&&{}  \notag \\
&&\Downarrow  \notag \\
&&{}  \notag \\
x^{\mu}_i(\tau ) &=& z^{\mu}_{wigner}(\tau , {\vec \eta}_i(\tau ))
= u^{\mu}(P)\, \tau + \epsilon^{\mu}_r(u(P))\, \eta^r_i(\tau ),
\notag \\
p^{\mu}_i(\tau ) &=& \sqrt{m^2_i + {\vec \kappa}_i^2(\tau )}\,
u^{\mu}(P) + \epsilon^{\mu}_r(u(P))\, \kappa_{ir}(\tau ).
 \label{25}
\end{eqnarray}

\medskip

While the potentials in $M_{(int)}$ determine ${\vec \rho}_q(\tau
)$ and ${ \vec \pi}_q(\tau )$ through Hamilton equations, the
potentials in ${\vec k} _{(int)}$ determine ${\vec q}_+({\vec
\rho}_q, {\vec \pi}_q)$. It is seen, therefore - as it should be
expected - that the relativistic theory of orbits is much more
complicated than in the non-relativistic case, where the absolute
orbits ${\vec \eta}_i(t)$ are proportional to the relative orbit
${ \vec \rho}_q(t)$ in the rest frame.

\vfill\eject

\section{A Simple 2-Particle Model with a-a-a-d Interaction.}

Instead of the physically more relevant but complicated system of
Ref.\cite{27}, whose internal Hamiltonian and boosts in the
rest-frame instant form are given in Eq.(\ref{9}), let us study a
simple two-body system with an a-a-a-d interaction, defined at the
end of Appendix A in terms of two first class constraints. As we
shall see its treatment in the constraint formalism  leads to a
realization of the Poincare' algebra only in the rest frame.
Therefore, let us look at its reformulation in the rest-frame
instant form, where the rest-frame conditions are automatically
contained.

\medskip

In the rest-frame instant form we may define the model by making
the ansatz that the generators of the internal realization of the
Poincare' algebra have the form [we use $\vec \rho = {\vec \eta}_1
- {\vec \eta}_2$ of Eqs.( \ref{19})]

\begin{eqnarray}
M_{(int)} &=&\sqrt{m_{1}^{2} + {\vec \kappa }_{1}^{2} + \Phi
({\vec \rho }^2)} + \sqrt{m_{2}^{2} + {\vec \kappa }_{2}^{2} +
\Phi ({\vec \rho }^2)},  \notag \\
\vec p &=& {\vec \kappa }_{1} + {\vec \kappa }_{2},  \notag \\
\vec j &=& {\vec \eta }_{1} \times {\vec \kappa}_{1} + {\vec \eta }_{2}
\times {\vec \kappa }_{2},  \notag \\
{\vec k}_{(int)} &=& - {\vec \eta }_{1}\, \sqrt{m_{1}^{2} + {\vec
\kappa } _{1}^{2} + \Phi ({\vec \rho }^2)} - {\vec \eta }_{2}\,
\sqrt{m_{2}^{2} + { \vec \kappa }_{2}^{2} + \Phi ({\vec \rho
}^2)}.
  \label{26}
\end{eqnarray}

\noindent Let us verify this ansatz by checking whether these generators
satisfy the Poincare' algebra.

\medskip

It is self evident that one has

\begin{equation}
\lbrack j_{i},j_{j}]=\varepsilon _{ijk}\,j_{k},\qquad \lbrack
j_{i},k_{(int)j}]=\varepsilon _{ijk}\,k_{(int)k}.
  \label{27}
\end{equation}

We examine the other Poincare' brackets. First note that

\begin{eqnarray}
\lbrack p_{i},k_{(int)j}] &=&[\kappa _{1i}+\kappa _{2i},-\eta
_{1j}\,\sqrt{ m_{1}^{2}+{\vec{\kappa}}_{1}^{2}+\Phi
({\vec{\rho}}^{2})}-\eta _{2j}\,\sqrt{
m_{2}^{2}+{\vec{\kappa}}_{2}^{2}+\Phi ({\vec{\rho}}^{2})}]=  \notag \\
&=&\delta _{ij}\,[\sqrt{m_{1}^{2}+{\vec{\kappa}}_{1}^{2}+\Phi
({\vec{\rho}} ^{2})}+\sqrt{m_{2}^{2}+{\vec{\kappa}}_{2}^{2}+\Phi
({\vec{\rho}}^{2})}]=\notag \\
&=&\delta _{ij}\,M_{(int)}.
  \label{28}
\end{eqnarray}

Next, examine [$\Phi^{^{\prime}}(x) = {\frac{{d\, \Phi(x)}}{{dx}}}$]

\begin{eqnarray}
\lbrack M_{(int)},k_{(int)i}]
&=&[\sqrt{m_{1}^{2}+{\vec{\kappa}}_{1}^{2}+\Phi (
{\vec{\rho}}^{2})}+\sqrt{m_{2}^{2}+{\vec{\kappa}}_{2}^{2}+\Phi
({\vec{\rho}}^{2})},  \notag \\
&&{}  \notag \\
&&-\eta _{1i}\,\sqrt{m_{1}^{2}+{\vec{\kappa}}_{1}^{2}+\Phi
({\vec{\rho}}^{2}) }-\eta
_{2i}\,\sqrt{m_{2}^{2}+{\vec{\kappa}}_{2}^{2}+\Phi
({\vec{\rho}}^{2})}
]=  \notag \\
&=&\kappa _{1i}+\kappa _{2i}-(\eta _{1i}+\eta
_{2i})\,[\sqrt{m_{1}^{2}+{\vec{ \kappa}}_{1}^{2}+\Phi
({\vec{\rho}}^{2})},\sqrt{m_{2}^{2}+{\vec{\kappa}}
_{2}^{2}+\Phi ({\vec{\rho}}^{2})}]=  \notag \\
&=&\kappa _{1i}+\kappa _{2i}-(\eta _{1i}+\eta
_{2i})\,[-\frac{2\,\Phi ^{\prime
}({\vec{\rho}}^{2})\,{\vec{\rho}}}{\sqrt{m_{1}^{2}+{\vec{\kappa}}
_{1}^{2}+\Phi ({\vec{\rho}}^{2})}}\cdot
\frac{{\vec{\kappa}}_{2}}{\sqrt{
m_{2}^{2}+{\vec{\kappa}}_{2}^{2}+\Phi ({\vec{\rho}}^{2})}}-  \notag \\
&-&\frac{{\vec{\kappa}}_{1}}{\sqrt{m_{1}^{2}+{\vec{\kappa}}_{1}^{2}+\Phi
({ \vec{\rho}}^{2})}}\cdot \frac{2\,\Phi ^{\prime
}({\vec{\rho}}^{2})\,{\vec{
\rho}}}{\sqrt{m_{2}^{2}+{\vec{\kappa}}_{2}^{2}+\Phi
({\vec{\rho}}^{2})}}]=
\notag \\
&=&\kappa _{1i}+\kappa _{2i}-(\eta _{1i}+\eta
_{2i})\,[-\frac{2\,\Phi ^{\prime
}({\vec{\eta}}^{2})\,{\vec{\rho}}}{\sqrt{m_{1}^{2}+{\vec{\kappa}}
_{1}^{2}+\Phi ({\vec{\rho}}^{2})}}\cdot
\frac{{\vec{\kappa}}_{1}+{\vec{\kappa
}}_{2}}{\sqrt{m_{2}^{2}+{\vec{\kappa}}_{2}^{2}+\Phi
({\vec{\rho}}^{2})}}].
\notag \\
&&{}  \label{29}
\end{eqnarray}

But the rest-frame condition $\vec p \approx 0$ implies

\begin{equation}
\lbrack M_{(int)},{\vec{k}}_{(int)}]\approx 0\approx \vec{p}.
\label{30}
\end{equation}

The remaining crucial bracket is

\begin{eqnarray}
\lbrack k_{(int)i},k_{(int)j}] &=&[-\eta
_{1i}\,\sqrt{m_{1}^{2}+{\vec{\kappa} }_{1}^{2}+\Phi
({\vec{\rho}}^{2})}-\eta _{2i}\,\sqrt{m_{2}^{2}+{\vec{\kappa}}
_{2}^{2}+\Phi ({\vec{\rho}}^{2})},  \notag \\
&&-\eta _{1j}\,\sqrt{m_{1}^{2}+{\vec{\kappa}}_{1}^{2}+\Phi
({\vec{\rho}}^{2}) }-\eta
_{2j}\,\sqrt{m_{2}^{2}+{\vec{\kappa}}_{2}^{2}+\Phi
({\vec{\rho}}^{2})}]=  \notag \\
&=&[\eta _{1i}\,\sqrt{m_{1}^{2}+{\vec{\kappa}}_{1}^{2}+\Phi
({\vec{\rho}} ^{2})},\eta
_{1j}\,\sqrt{m_{1}^{2}+{\vec{\kappa}}_{1}^{2}+\Phi ({\vec{\rho}}
^{2})}]+  \notag \\
&+&[\eta _{2i}\,\sqrt{m_{2}^{2}+{\vec{\kappa}}_{2}^{2}+\Phi
({\vec{\rho}} ^{2})},\eta
_{2j}\,\sqrt{m_{2}^{2}+{\vec{\kappa}}_{2}^{2}+\Phi ({\vec{\rho}}
^{2})}]+  \notag \\
&+&[\eta _{1i}\,\sqrt{m_{1}^{2}+{\vec{\kappa}}_{1}^{2}+\Phi
({\vec{\rho}} ^{2})},\eta
_{2j}\,\sqrt{m_{2}^{2}+{\vec{\kappa}}_{2}^{2}+\Phi ({\vec{\rho}}
^{2})}]+  \notag \\
&+&[\eta _{2i}\,\sqrt{m_{2}^{2}+{\vec{\kappa}}_{2}^{2}+\Phi
({\vec{\rho}} ^{2})},\eta
_{1j}\,\sqrt{m_{1}^{2}+{\vec{\kappa}}_{1}^{2}+\Phi ({\vec{\rho}}
^{2})}]=  \notag \\
&=&\eta _{1j}\,\kappa _{1i}-\eta _{1i}\,\kappa _{1j}+\eta _{2j}\,\kappa
_{2i}-\eta _{2i}\,\kappa _{2j}-  \notag \\
&-&\eta _{1i}\,\eta _{2j}\,\frac{2\,\Phi ^{\prime
}({\vec{\rho}}^{2})\,{\vec{ \rho}}\cdot
({\vec{\kappa}}_{1}+{\vec{\kappa}}_{2})}{\sqrt{m_{1}^{2}+{\vec{
\kappa}}_{1}^{2}+\Phi
({\vec{\rho}}^{2})}\,\sqrt{m_{2}^{2}+{\vec{\kappa}}
_{2}^{2}+\Phi ({\vec{\rho}}^{2})}}+  \notag \\
&+&\eta _{2i}\,\eta _{1j}\,\frac{2\,\Phi ^{\prime
}({\vec{\rho}}^{2})\,{\vec{ \rho}}\cdot
({\vec{\kappa}}_{1}+{\vec{\kappa}}_{2})}{\sqrt{m_{1}^{2}+{\vec{
\kappa}}_{1}^{2}+\Phi
({\vec{\rho}}^{2})}\,\sqrt{m_{2}^{2}+{\vec{\kappa}} _{2}^{2}+\Phi
({\vec{\rho}}^{2})}}.
  \label{31}
\end{eqnarray}

Again the rest-frame condition implies

\begin{eqnarray}
\lbrack k_{(int)i},k_{(int)j}] &\approx &\eta _{1j}\,\kappa _{1i}-\eta
_{1i}\,\kappa _{1j}+\eta _{2j}\,\kappa _{2i}-\eta _{2i}\,\kappa _{2j}=
\notag \\
&=&-\varepsilon _{ijk}\,\varepsilon _{klm}\,(\eta _{1l}\,\kappa _{1m}+\eta
_{2l}\,\kappa _{2m})=-\varepsilon _{ijk}\,j_{k}.
  \label{32}
\end{eqnarray}

\bigskip

From Eq.(\ref{10}) the interacting form of the canonical internal 3-center
of mass is weakly equal to the 3-center of energy due to the rest frame
condition $\vec p \approx 0$

\begin{equation}
{\vec q}_+ \approx {\vec R}_+ = - \frac{{\vec
k}_{(int)}}{M_{(int)}},
 \label{33}
\end{equation}

Eqs.(\ref{19}), (\ref{20}) (as well as Eq.(\ref{90}) in Appendix
A) imply the following forms of ${\vec{k}}_{(int)}$ and
$M_{(int)}$

\begin{eqnarray}
{\vec{k}}_{(int)} &\approx
&-{\vec{\eta}}_{1}\,\sqrt{m_{1}^{2}+{\vec{\pi}} ^{2}+\Phi
({\vec{\rho}}^{2})}-{\vec{\eta}}_{2}\,\sqrt{m_{2}^{2}+{\vec{\pi}}
^{2}+\Phi ({\vec{\rho}}^{2})},  \notag \\
M_{(int)} &\approx &\sqrt{m_{1}^{2}+{\vec{\pi}}^{2}+\Phi
({\vec{\rho}}^{2})}+ \sqrt{m_{2}^{2}+{\vec{\pi}}^{2}+\Phi
({\vec{\rho}}^{2})}=\mathcal{M}_{(int)}.
 \label{34}
\end{eqnarray}
In the free case, with
$\mathcal{M=}\sqrt{m_{1}^{2}+{\vec{\pi}}^{2}}+\sqrt{
m_{2}^{2}+{\vec{\pi}}^{2}},~$Eqs.(\ref{20}) imply

\begin{eqnarray}
{\vec \eta }_{1} &\approx &{\vec q}_+ + \frac{1}{2}\, (1 - \frac{
m_{1}^{2}-m_{2}^{2}}{\mathcal{M}^{2}})\, \vec \rho \approx
\frac{1}{2}\, (1
- \frac{m_{1}^{2}-m_{2}^{2}}{\mathcal{M}^{2}})\, \vec \rho  \notag \\
{\vec \eta }_{2} &\approx &{\vec q}_+ - \frac{1}{2}\, (1 + \frac{
m_{1}^{2}-m_{2}^{2}}{\mathcal{M}^{2}})\, \vec \rho \approx -
\frac{1}{2}\, (1 + \frac{m_{1}^{2}-m_{2}^{2}}{\mathcal{M}^{2}})\,
\vec \rho.
  \label{35}
\end{eqnarray}

\noindent and also

\begin{eqnarray}
\sqrt{m_{1}^{2} + {\vec \pi }^{2}} &=&\frac{1}{2}\, (\mathcal{M} +
\Delta )\, = \frac{\mathcal{M}}{2}\, (1 + \frac{
m_{1}^{2}-m_{2}^{2}}{\mathcal{M}^{2}}),  \notag \\
\sqrt{m_{2}^{2} + {\vec \pi }^{2}} &=&\frac{1}{2}\, (\mathcal{M} - \Delta )
= \frac{\mathcal{M}}{2}\, (1 - \frac{ m_{1}^{2}-m_{2}^{2}}{\mathcal{M}^{2}}),
\label{36}
\end{eqnarray}

\noindent where $\Delta
=\sqrt{m_{1}^{2}+{\vec{\pi}}^{2}}-\sqrt{m_{2}^{2}-{
\vec{\pi}}^{2}}$, $\mathcal{M}\,\Delta =m_{1}^{2}-m_{2}^{2}$.

Therefore in the free case we have the following expression for
the 3-coordinates ${\vec \eta}_i$

\begin{eqnarray}
{\vec \eta }_{1} &\approx & {\vec q}_+ + {1\over 2}\, (1 - {{m^2_1
- m^2_2}\over {{\cal M}^2}})\, \vec \rho = {\vec q}_+ +
{\frac{\sqrt{m_2^2 + {\vec \pi}^2}}{{
\mathcal{M}}}}\, \vec \rho,  \notag \\
{\vec \eta }_{2} &\approx & {\vec q}_+ - {1\over 2}\, (1 + {{m^2_1
- m^2_2}\over {{\cal M}^2}})\, \vec \rho = {\vec q}_+ -
{\frac{\sqrt{m_1^2 + {\vec \pi}^2} }{ {\mathcal{M}}}}\, \vec \rho.
\label{37}
\end{eqnarray}

\bigskip

If we use the canonical basis ${\vec q}_+$, $\vec p \approx 0$,
$\vec \rho$, $\vec \pi$ of the free case also in our simple
interacting case, Eqs.(\ref{37}) must be replaced by
Eqs.(\ref{25}). To this end we must find the functions ${\vec
q}_+(\vec \rho ,\vec \pi )$ from the gauge conditions ${\vec q}
_+^{(int)} \approx {\vec R}^{(int)}_+ \approx 0$, namely from
${\vec k}_{(int)} \approx 0 $.

\medskip

In our simple interacting case, from Eq. (\ref{19}), (\ref{34})
and (\ref{37}) we get

\begin{eqnarray}
- \mathcal{M}_{(int)}\,\, {\vec q}_+^{(int)} &\approx& {\vec
k}_{(int)} \approx
\notag \\
&\approx& - {\vec q}_+\, \Big[\sqrt{m_{1}^{2} + {\vec \pi }^{2} +
\Phi ( {\vec \rho}^2)}\, + \sqrt{m_{2}^{2} + {\vec \pi }^{2} +
\Phi ({\vec \rho }^2)}\Big] -
\notag \\
&-& \vec \rho\, \frac{\sqrt{m_{2}^{2} + {\vec \pi }^{2}}\, \sqrt{
m_{1}^{2} + {\vec \pi }^{2} + \Phi ({\vec \rho}^2)} -
\sqrt{m_{1}^{2} + {\vec \pi }^{2} }\, \sqrt{m_{2}^{2} + {\vec \pi
}^{2} + \Phi ({\vec \rho }^2)}}{\mathcal{M}}.
\notag \\
&&{}  \label{38}
\end{eqnarray}

\medskip

If, in analogy to the free case, we define $\Delta_{(int)} =
\sqrt{m_{1}^{2} + {\vec \pi}^2 +\Phi ( {\vec \rho }^2)} -
\sqrt{m_{2}^{2} + {\vec \pi}^2 + \Phi ({\vec \rho }^2)}$, we have
$\mathcal{M}_{(int)}\, \Delta_{(int)} = m^2_1 - m_2^2$ and we get

\begin{eqnarray}
\sqrt{m_{1}^{2} + {\vec \pi}^2 +\Phi ( {\vec \rho }^2)} &=&
{\frac{1}{2}}\, ( \mathcal{M}_{(int)} + \Delta_{(int)}) =
{\frac{{\mathcal{M}_{(int)}}}{2}}\, (1 + {
\frac{{m^2_1 - m^2_2}}{{\mathcal{M}_{(int)}^2}}}),  \notag \\
\sqrt{m_{2}^{2} + {\vec \pi}^2 +\Phi ( {\vec \rho }^2)} &=&
{\frac{1}{2}}\, ( \mathcal{M}_{(int)} - \Delta_{(int)}) =
{\frac{{\mathcal{M}_{(int)}}}{2}}\, (1 - {
\frac{{m^2_1 - m^2_2}}{{\mathcal{M}_{(int)}^2}}}).  \notag \\
&&{}  \notag \\
&&\Downarrow  \notag \\
&&{}  \notag \\
\mathcal{M} &=& \sum_{i=1}^2\, \sqrt{m_i^2 + {\vec \pi}^2} =
\sqrt{\Big[{\frac{{\mathcal{M}_{(int)}}}{2}}\, (1 + {\frac{{ m^2_1
- m^2_2}}{{\mathcal{M}_{(int)}^2}}})\Big]^2 - \Phi({\vec \rho}^2)}
+ \notag \\
&+& \sqrt{\Big[{\frac{{\mathcal{M}_{(int)}}}{2}}\, (1 -
{\frac{{m^2_1 - m^2_2} }{{\mathcal{M}_{(int)}^2}}})\Big]^2 -
\Phi({\vec \rho}^2)}.
 \label{39}
\end{eqnarray}

\medskip

As a consequence, Eq.(\ref{38}) may be written in the form

\begin{eqnarray}
{\vec{k}}_{(int)} &\approx &-\mathcal{M}_{(int)}\,\, \Big({\vec{q}}_{+}+  \notag \\
&+&\vec{\rho}\,\frac{(\sqrt{m_{2}^{2}+{\vec{\pi}}^{2}}-\sqrt{m_{1}^{2}+{\vec{
\pi}}^{2}})+(\sqrt{m_{2}^{2}+{\vec{\pi}}^{2}}+\sqrt{m_{1}^{2}+{\vec{\pi}}^{2}
})\,\frac{m_{1}^{2}-m_{2}^{2}}{\mathcal{M}_{(int)}^{2}}}{2\,\mathcal{M}}\Big)
= \notag \\
&=&-\mathcal{M}_{(int)}\,{\vec{q}}_{+}+\vec{\rho}\,\mathcal{M}_{(int)}\,\frac{
m_{1}^{2}-m_{2}^{2}}{2\,\mathcal{M}^{2}}-\vec{\rho}\,\frac{
m_{1}^{2}-m_{2}^{2}}{2\,\mathcal{M}_{(int)}}=  \notag \\
&=&-\mathcal{M}_{(int)}\,{\vec{q}}_{+}+\frac{m_{1}^{2}-m_{2}^{2}}{2}\,(\frac{
\mathcal{M}_{(int)}}{\mathcal{M}^{2}}-\frac{1}{\mathcal{M}_{(int)}})\,\vec{\rho}.
\label{40}
\end{eqnarray}

Therefore the gauge fixing condition ${\vec q}_+^{(int)} \approx
0$, or ${\vec k}_{(int)} \approx 0$, gives

\begin{equation}
{\vec{q}}_{+}\approx {\vec{q}}_{+}(\vec{\rho},\vec{\pi})=\frac{
m_{1}^{2}-m_{2}^{2}}{2}\,(\frac{1}{\mathcal{M}^{2}}-\frac{1}{\mathcal{M}
_{(int)}^{2}})\,\vec{\rho},
  \label{41}
\end{equation}

\noindent so that, by using the inverse canonical transformation
(\ref{20}), in our simple interacting case we get the following
reconstruction of the 3-coordinates ${\vec \eta}_i$ and of the
4-coordinates $x^{\mu}_i$

\begin{eqnarray}
{\vec{\eta}}_{1} &\approx &{\frac{1}{2}}\,[1-
{{m_{1}^{2}-m_{2}^{2}}\over{
\mathcal{M}_{(int)}^{2}}}]\,\vec{\rho} \,\,  {\rightarrow
_{c\rightarrow \infty }} \,\,
\frac{m_{2}}{m_{1}+m_{2}}\,\vec{\rho}
,  \notag \\
{\vec{\eta}}_{2} &\approx
&-{\frac{1}{2}}\,[1+{{m_{1}^{2}-m_{2}^{2}}\over{
\mathcal{M}_{(int)}^{2}}}]\,\vec{\rho}  \notag \\
{\rightarrow _{c\rightarrow \infty }}
&&-\frac{m_{1}}{m_{1}+m_{2}}\,\vec{\rho },\nonumber \\
 &&{}\nonumber \\
 x^{\mu}_i(\tau ) &=& u^{\mu}(P)\, \tau + {1\over 2}\,
\Big[(-1)^{i+1} - {{m^2_1 - m^2_2}\over {\sum_{j=1}^2\,
\sqrt{m^2_j + {\vec \pi}^2 + \Phi ({\vec \rho}^2)}}}\Big]\,
\epsilon^{\mu}_r(P)\, \rho^r.
  \label{42}
\end{eqnarray}

This completes the reconstruction of the relativistic orbits of
the two particles.
\medskip

We see that in the non-relativistic limit we recover the standard
result in the center of mass frame $\vec{p}=0$. This is equivalent
to adding the first class constraints $\vec{p}\approx 0$ to the
non-relativistic Hamiltonian $H_{com}$ and to fix the gauge by
putting the non-relativistic center of mass at the origin

\begin{equation}
\vec x = \frac{m_{1}\, {\vec \eta }_{1} + m_{2}\, {\vec \eta }_{2}}{m_{1} +
m_{2}} \approx 0.  \label{43}
\end{equation}

\bigskip

For our simple interacting relativistic model we get that the
rest-frame 3-coordinates ${\vec \eta}_i$ are still proportional to
the relative variable $\vec \rho$ [for more complex models there
could be a component along $\vec \pi$ coming from the function
${\vec q}_+(\vec \rho , \vec \pi)$]. However, instead of the
numerical proportionality constants of the non-relativistic case,
we have a non-trivial dependence \textit{ on the total constant
fixed c.m. energy} ($\mathcal{M}_{(int)} = \sum_{i=1}^2\,
\sqrt{m^2_i + {\vec \pi}^2 + \Phi ({\vec \rho}^2)}$).

\vfill\eject

\section{Evaluation of the Orbits in the Simple Relativistic Two-Body
Problem with a Coulomb-like Potential.}

For illustrative purposes, we make the following choice for the a-a-a-d
potential $\Phi$

\begin{equation}
\Phi ({\vec \rho}^2) = - 2\, \mu\, \frac{e^{2}}{\rho}, \qquad \rho
= \sqrt{{ \vec \rho}^2}.
  \label{44}
\end{equation}

This Coulomb-like potential is not to be confused with the real Coulomb
potential between charged particles, which is outside the square roots as
shown in Eq.(\ref{9}) and which produces completely different relativistic
orbits. However both models may have the same non-relativistic limit for
suitable choices of the parameters

\bigskip

The invariant mass $M_{(int)}$ of the two-body model (the
Hamiltonian of its relative motion) in the rest-frame instant form
is

\begin{equation}
M_{(int)} \approx \mathcal{M}_{(int)} = \sqrt{m_{1}^{2} + {\vec
\pi}^2 - 2\, \mu\, \frac{e^{2}}{\rho }} + \sqrt{m_{2}^{2} + {\vec
\pi}^2 - 2\, \mu\, \frac{e^{2}}{\rho }}.
 \label{45}
\end{equation}

Instead of studying the Hamilton equations for $\vec \rho$, $\vec
\pi$ with $ \mathcal{M}_{(int)}$ as Hamiltonian, we will find the
orbits using Hamilton-Jacobi methods \footnote{ See Refs.\cite{33}
for the relativistic Kepler or Coulomb problem with respect to a
fixed center and Refs.\cite{34} for its use. Let us remark that
the techniques of Refs.\cite{33} could be applied to the results
of Ref.\cite{27} to describe the relativistic Darwin two-body
problem. Instead it is not known the internal boost $\vec k$,
satisfying $\{ p_i, k_j \} = \delta_{ij}\, M$, in the case of the
pure Coulomb interaction: $M =\sum_{i=1}^2\, \sqrt{m^2_i + {\vec
\kappa}^2} + {{Q_1\, Q_2}\over {4\pi\, |{\vec \eta}_1 - {\vec
\eta}_2|}}$}. Since the potential is a central one, our orbit is
confined to a plane with

\begin{equation}
{\vec \pi }^{2}=\pi _{\rho }^{2} + \frac{\pi _{\phi }^{2}}{\rho ^{2} }.
\label{46}
\end{equation}

Since both the time and the angle are cyclic, the generating function is

\begin{equation}
S = W_{1}(\rho ) + \alpha _{\phi }\, \phi - w\, t \equiv W_{1}(\rho ,\phi )
- w\, t,
  \label{47}
\end{equation}

\noindent with $w$ the invariant total c.m. energy. \ The Hamilton-Jacobi
equation is

\begin{equation}
\sqrt{(\frac{\partial W_{1}}{\partial \rho })^{2} + \frac{\alpha
_{\phi }^{2} }{ \rho ^{2}} + m_{1}^{2} - 2\, \mu\,
\frac{e^{2}}{\rho }} + \sqrt{(\frac{
\partial W_{1}}{ \partial \rho })^{2} + \frac{\alpha _{\phi }^{2}}{\rho ^{2}}
+ m_{2}^{2} - 2\, \mu\, \frac{e^{2}}{\rho }} = w,
  \label{48}
\end{equation}

\noindent This leads to [the function $b^{2}(w)$ is defined in Eq.(\ref{83})
of Appendix A]

\begin{equation}
(\frac{\partial W_{1}}{\partial \rho })^{2} + \frac{\alpha _{\phi
}^{2}}{ \rho ^{2}} - 2\, \mu\, \frac{e^{2}}{\rho } = b^{2}(w),
\label{49}
\end{equation}

\noindent and so we get

\begin{equation}
W_{1}(\rho ,\phi )=\int d\rho\, \sqrt{b^{2}(w) - \frac{\alpha
_{\phi }^{2}}{ \rho ^{2}} + 2\, \mu\, \frac{e^{2}}{\rho }} +
\alpha _{\phi }\, \phi .
 \label{50}
\end{equation}

The new coordinate canonically conjugate to the new momentum $\alpha _{\phi
} $ is the constant

\begin{equation}
\beta _{2} = \frac{\partial W}{\partial \alpha _{\phi }} = -
\int\, \frac{ \alpha _{\phi }\, d\rho }{\rho ^{2}\, \sqrt{b^{2}(w)
- \frac{\alpha _{\phi }^{2}}{\rho ^{2}} + 2\, \mu\,
\frac{e^{2}}{\rho }}} + \phi .
  \label{51}
\end{equation}

If we define

\begin{equation}
u = \frac{1}{\rho },
  \label{52}
\end{equation}

\noindent we get

\begin{equation}
\beta _{2} = \frac{\partial W}{\partial \alpha _{\phi }} = \int\,
\frac{ \alpha _{\phi }\, du}{\sqrt{b^{2}(w) - \alpha _{\phi
}^{2}\, u^{2} + 2\, \mu\, e^{2}\, u}} + \phi ,
  \label{53}
\end{equation}

\noindent or

\begin{equation}
\phi = \beta _{2} = \int\, \frac{du}{\sqrt{\frac{b^{2}(w)}{\alpha _{\phi
}^{2}} + \frac{2\, \mu\, e^{2}\, u}{\alpha _{\phi }^{2}} - u^{2}}}.
\label{54}
\end{equation}

\noindent This result leads to the ellipse (we consider only
bounded orbits)

\begin{equation}
\frac{1}{\rho }=\frac{\mu \,e^{2}}{\alpha _{\phi
}^{2}}\,[1+\sqrt{1+\frac{ b^{2}(w)\,\alpha _{\phi }^{2}}{\mu
^{2}\,e^{2}}}\,\cos (\phi -\beta _{2})].
 \label{55}
\end{equation}
Eqs.(\ref{54}) and (\ref{55}) allow us to determine the orbit of the
relative motion

\begin{equation}
\vec \rho = \rho\, (\cos \phi\, \mathbf{i} + \sin \phi\, \mathbf{j)}.
\label{56}
\end{equation}

Let us compare these results with the non-relativistic limit. In
the non-relativistic Kepler or Coulomb case \cite{1},
Eq.(\ref{55}) is replaced by the following expression (in the
non-relativistic limit we have $b^2(w) \rightarrow 2\, \mu\, E$
with the energy $E = w - m_1 - m_2$)

\begin{equation}
\frac{1}{\rho }=\frac{\mu \,e^{2}}{\alpha _{\phi
}^{2}}\,[1+\sqrt{1+\frac{ 2\,E\,\alpha _{\phi }^{2}}{\mu
\,e^{2}}}\,\cos (\phi -\beta _{2})].
 \label{57}
\end{equation}
If we use the non-relativistic limit into Eqs.(\ref{42}) for the relation
among ${\vec{\eta}}_{i}$ and $\vec{\rho}$, we get the non-relativistic
expressions

\begin{eqnarray}
{\vec \eta }_{1} &=& \frac{\alpha _{\phi }^{2}}{m_{1}\, e^{2}}\, \frac{1}{[1
+ \sqrt{1 + \frac{2\, E\, \alpha _{\phi }^{2}}{\mu\, e^{2}}}\, \cos (\phi
-\beta _{2})]}\, (\cos \phi\, \mathbf{i} + \sin \phi\, \mathbf{j)},  \notag
\\
{\vec \eta }_{2} &=& - \frac{\alpha _{\phi}^{2}}{m_{2}\, e^{2}}\,
\frac{1}{ [1 + \sqrt{1 + \frac{2\, E\, \alpha _{\phi }^{2}}{\mu\,
e^{2}}}\, \cos (\phi -\beta _{2})]}\, (\cos \phi\, \mathbf{i} +
\sin \phi\, \mathbf{j).}
 \label{58}
\end{eqnarray}

\bigskip

For the relativistic counterparts, given in Eq. (\ref{42}), we
have from Eqs.( \ref{39})

\begin{eqnarray}
w &=& \mathcal{M}_{(int)} = \sum_{i=1}^2\, \sqrt{m^2_i + {\vec
\pi}^2 - 2 \mu\, {{e^2}\over {\rho}}},  \notag \\
&&{}  \notag \\
\mathcal{M} &=& \sqrt{\Big[{\frac{{w}}{2}}\, (1 + {\frac{{m^2_1 -
m^2_2}}{{
w^2}}})\Big]^2 + 2\, \mu\, {\frac{{e^2}}{{\rho}}}} +  \notag \\
&+& \sqrt{\Big[{\frac{{w}}{2}}\, (1 - {\frac{{m^2_1 -
m^2_2}}{{w^2}}})\Big] ^2 + 2\, \mu\, {\frac{{e^2}}{{\rho}}}}
\equiv \mathcal{M}(w,\rho ).
 \label{59}
\end{eqnarray}

In this case, from Eqs.(\ref{42}) we have

\begin{eqnarray}
{\vec{\eta}}_{1} &\approx &{\frac{1}{2}}\,\rho \,(\cos \phi
\,\mathbf{i} +\sin \phi
\,\mathbf{j})\,[1-{{m_{1}^{2}-m_{2}^{2}}\over{w^{2}}}],
\notag \\
{\vec{\eta}}_{2} &\approx &-{\frac{1}{2}}\,\rho \,(\cos \phi
\,\mathbf{i} +\sin \phi
\,\mathbf{j})\,[1+{{m_{1}^{2}-m_{2}^{2}}\over{w^{2}}}],
 \label{60}
\end{eqnarray}

\noindent where for $\rho $ we have to use the solution given in
Eq.(\ref{55}).
\bigskip

We have the following situation:

\noindent 1) For equal masses, the relativistic and
non-relativistic expressions are identical with $b^2({\cal
M}_{(int)}) \mapsto 2\, \mu\, E$.
\medskip

\noindent 2) In the limit in which one of the masses becomes very
great (say $m_{2}$) then, since we have

\begin{equation}
\frac{m_{1}^{2}-m_{2}^{2}}{w^{2}}\, \rightarrow_{m_2 \rightarrow
\infty}\,\, -1,
 \label{61}
\end{equation}

\noindent the relativistic and non-relativistic expressions are identical
also. \medskip

\noindent 3) If we introduce the new notation

\begin{equation}
w=\frac{m_{1}+m_{2}}{\sqrt{\omega }},
  \label{62}
\end{equation}

\noindent then the relativistic orbits become

\begin{eqnarray}
{\vec{\eta}}_{1} &=&{\frac{1}{2}}\,\rho (\phi )\,(\cos \phi
\,\mathbf{i} +\sin \phi
\,\mathbf{j)}\,[1-\frac{m_{1}-m_{2}}{m_{1}+m_{2}}\,\omega ],
\notag \\
{\vec{\eta}}_{2} &=&-{\frac{1}{2}}\,\rho (\phi )\,(\cos \phi
\,\mathbf{i} +\sin \phi
\,\mathbf{j)}\,[1+\frac{m_{1}-m_{2}}{m_{1}+m_{2}}\omega ].
\label{63}
\end{eqnarray}

\bigskip

Since $\omega $ is a constant of motion, the main difference
between the relativistic and the non-relativistic orbits is the
proportionality constant between individual particle coordinates
and the relative coordinate, which, however, is now dependent on
the invariant mass of the system.

\vfill\eject

\section{Conclusions}

We have given a complete treatment of the Hamiltonian two-body
problem in the rest-frame instant form, arising from parametrized
Minkowski theories when the dynamics is described with respect to
the inertial intrinsic rest frame of the isolated system with its
simultaneity 3-surfaces given by the Wigner hyper-planes. The
existence of two realizations of the Poincare' group (the external
one and the unfaithful internal one inside the Wigner
hyper-planes), together with the clarification of  the only three
intrinsic notions of a center-of-mass-like collective variable
allow us to solve all the kinematical problems and to define
canonical transformations for the separation of the center of mass
from the relative motion as is possible in Newtonian mechanics.

\bigskip

In the rest-frame instant form of dynamics there is a natural
gauge fixing ${ \vec{k}}_{(int)} \approx 0$ to the rest-frame
conditions $\vec{p}\approx 0$, which allows us to clarify
completely the determination of the relativistic orbits inside the
Wigner hyper-planes. With this gauge fixing it is possible to
describe the isolated system from the point of view of an inertial
observer, whose world-line $x_{s}^{\mu }(\tau )=Y^{\mu }(\tau
)=u^{\mu }(P)\,\tau $ is the (covariant non-canonical)
Fokker-Pryce center of inertia and the only non-vanishing
generators of the internal Poincare' algebra are the
interaction-dependent invariant mass $M$ and the
interaction-independent spin ${\vec {\bar S}}$. For every isolated
system there is the universal realization (\ref{6}) of the
external Poincare' algebra, whose generators depend upon the
external canonical 3-center-of-mass variables $\vec z$, $\vec h$
\footnote{See footnote 9. Even if $\vec z$ is not covariant due to
the no-interaction theorem, there is no problem, because it
describes a globally decoupled pseudo-particle: all physical
results are described by the Wigner-covariant relative variables
inside the Wigner hyper-planes.} and upon the system, but only
through its invariant mass $M$ and its spin ${\vec {\bar S}}$. The
simplest model with a-a-a-d interaction is studied in detail.

\bigskip

To reconstruct the actual trajectories in Minkowski space-time in the above
inertial frame, we have to use Eqs.(\ref{4})

\begin{eqnarray}
x^{\mu}_i(\tau ) &=& u^{\mu}(P)\, \tau + \epsilon^{\mu}_r(P)\, \eta_i(\tau ),
\notag \\
p^{\mu}_i(\tau ) &=& \sqrt{m^2_i + {\vec \kappa}_i^2(\tau )}\, u^{\mu}(P) +
\epsilon^{\mu}_r(P)\, \kappa_{ir}(\tau ).
  \label{64}
\end{eqnarray}

To eliminate the momenta and to get a purely configurational
description one should invert the first half of the Hamilton
equations, ${\dot {\vec \rho}} = \{ \vec \rho ,
\mathcal{M}_{(int)} \}$, to get $\vec \pi$ in terms of $\vec \rho$
and ${\dot {\vec \rho}}$ ($\dot f = {\frac{{df}}{{d\tau}}}$).

\medskip

Under Lorentz transformations $\Lambda$ generated by the external
Poincare' group, under which we have $\epsilon^{\mu}_r(u(\Lambda
P))=(R^{-1}(\Lambda ,P))_r{}^s\, \Lambda^{\mu}{}_{\nu}\,
\epsilon^{\nu}_s(u(P))$ and $\eta^{{^{\prime}} r}_i =$\break
$R^r{}_s(\Lambda ,P)\, \eta^s_i$ (see footnote 6), the derived
quantities $x^{\mu}_i$ and $p^{\mu}_i$ transform covariantly as
4-vectors \footnote{ Namely the rest-frame instant form satisfies
the {\it world-line condition}, since its synchronization of
clocks (the one-to-one correlation between the world-lines) is a
generalization of the gauge fixing $P \cdot (x_1 - x_2) \approx 0$
in models with second class constraints, as shown in Ref.\cite{13}
. As a consequence the world-lines have an objective existence.
However, in parametrized Minkowski theories one could choose
different 3+1 splittings of Minkowski space-times corresponding to
different one-to-one correlations (different conventions for the
synchronization of clocks). Since each 3+1 splitting is equivalent
to a different choice of non-inertial frame \cite{20} with its
inertial forces (see Appendix B for the non-inertial rest frames),
the new world-lines will be different (they are obtained from
those in the rest-frame instant form by means of a gauge
transformation sending an inertial frame into a non-inertial one
\cite{21,23} . This is the interpretation of the so-called {\it
frame dependence of the world-lines} quoted in Ref.\cite{13},
where it was connected to a semantic problem.}. However the
world-lines $x^{\mu}_i(\tau )$ are no more canonical variables,
because they depend on the (non-canonical) Fokker-Pryce center of
inertia. This, together with the non-covariance of the canonical
center of mass ${\tilde x} ^{\mu}$, is the way out from the
no-interaction theorem in the rest-frame instant form. \medskip

Once the world-lines $x^{\mu}_i(\tau )$ are known in terms of the
rest-frame time $\tau = T_s = u(P) \cdot x_s = u(P) \cdot Y = u(P)
\cdot \tilde x$, one can geometrically introduce different affine
parameters $\tau_i$ on each world-line. Then as shown in
Ref.\cite{13} one could arrive at the result $ x^{\mu}_i(\tau ) =
{\hat x}_i^{\mu}(\tau_1, \tau_2) = q^{\mu}(\tau_i)$, where
$q^{\mu}_i(\tau_i)$ are the Droz-Vincent predictive covariant
coordinates quoted in point b) of Section II and obtained from the
equations given in Appendix A by taking $x^{\mu}_i = q^{\mu}_i$ as
a Cauchy condition on a Wigner hyper-plane. These predictive
coordinates implement the geometrical idea that each world-line
may be reparametrized independently from the other even in
presence of interactions.

\bigskip

Having understood both the kinematical and dynamical problems of
relativistic orbit theory, the next step is to try to define a
perturbation theory around relativistic orbits as has been done in
the non-relativistic case \cite{1}: it could be relevant for the
special relativistic approximation of relativistic binaries in
general relativity, till now treated only in the post-Newtonian
approximation \cite{35}.

\vfill\eject

\appendix

\section{Two-Body Relativistic Hamiltonian Mechanics with Two First-Class
Constraints.}

In constraint dynamics for classical relativistic spinless
particles one begins by introducing compatible generalized mass
shell constraints. \ We work with constraints that involve
potentials that are independent of the relative momenta ($P^{\mu}
= p_1^{\mu} + p_2^{\mu}$, $M = \sqrt{P^2}$, $ r^{\mu} = x_1^{\mu}
- x_2^{\mu}$, $\{ x^{\mu}_i, p^{\nu}_j \} = - \delta_{ij}\,
\eta^{\mu\nu}$)

\begin{eqnarray}
\mathcal{H}_{1} &=&p_{1}^{2}-m_{1}^{2}-\Phi _{1}(r,P)\approx 0,  \notag \\
\mathcal{H}_{2} &=&p_{2}^{2}-m_{2}^{2}-\Phi _{2}(r,P)\approx 0.
  \label{65}
\end{eqnarray}

We call the scalars $\Phi _{i}(r,P)$ quasi-potentials (energy
dependent potentials that describe deviations from the free mass
shell conditions $ p_{i}^{2}-m_{i}^{2}\approx 0)$. Assuming that
the model derives from an unknown reparametrization invariant
Lagrangian (so that the canonical Hamiltonian vanishes), the
Hamiltonian is defined in terms of the constraints only by

\begin{equation}
\mathcal{H=\lambda }_{1}(\tau )\, \mathcal{H}_{1} + \lambda _{2}(\tau )\,
\mathcal{H}_{2}.
  \label{66}
\end{equation}

\noindent in which $\lambda _{i}$ are arbitrary Lagrange
multipliers (called Dirac multipliers). \ The constraints forming
this Hamiltonian must be such  that the time rate of change of the
constraints vanishes when the constraint are imposed. \ With the
time rate of change of an arbitrary dynamical variable $f$ given
by

\begin{equation}
\frac{df}{d\tau }=\{f,\mathcal{H\}},
  \label{67}
\end{equation}

\noindent we get \footnote{The Dirac multipliers, being functions
only of $\tau$, have zero Poisson bracket with phase space
functions.}

\begin{eqnarray}
\frac{d\mathcal{H}_{1}}{d\tau } &=& \{ \mathcal{H}_{1}, \mathcal{H}\} =
\lambda _{1}(\tau )\, \{\mathcal{H}_{1}, \mathcal{H}_{1}\} + \lambda
_{2}(\tau )\, \{\mathcal{H}_{1}, \mathcal{H}_{2}\} \approx  \notag \\
&\approx & \lambda_{2}(\tau )\, \{\mathcal{H}_{1}, \mathcal{H}_{2}\},
 \label{68}
\end{eqnarray}

\noindent and similarly

\begin{equation}
\frac{d\mathcal{H}_{2}}{d\tau }\approx \lambda _{1}(\tau )\,
\{\mathcal{H} _{2}, \mathcal{H}_{1}\}.
  \label{69}
\end{equation}

Thus, the constraints are constants (their time derivatives weakly
vanish) provided that

\begin{equation}
\{\mathcal{H}_{1},\mathcal{H}_{2}\}\approx 0.
  \label{70}
\end{equation}
This is called the compatibility condition. \ Thus, we must have

\begin{equation}
2\, p_{1} \cdot \{p_{1}, \Phi _{2}\} + 2\, p_{2}\cdot \{\Phi _{1}, p_{2}\} +
\{\Phi _{1}, \Phi _{2}\} \approx 0.
  \label{71}
\end{equation}

We assume that the scalar functions $\Phi_i$ depend on the following
variables

\begin{equation}
\Phi _{i} = \Phi _{i}(\frac{r_{\perp }^{2}}{2},
\frac{r_{||}^{2}}{2},M),
 \label{72}
\end{equation}

\noindent where

\begin{equation}
r_{||}^{\mu }=\frac{r\cdot P}{w^{2}}\,P^{\mu },\qquad r_{\perp }^{\mu
}=r^{\mu }-r_{||}^{\mu },\qquad P\cdot r_{\perp }=0.
  \label{73}
\end{equation}

Thus, our compatibility condition becomes

\begin{eqnarray}
&&-4\,p_{1}\cdot r_{\perp }\,\frac{\partial \Phi _{2}}{\partial
r_{\perp }^{2}}-4\,p_{1}\cdot r_{||}\,\frac{\partial \Phi
_{2}}{\partial r_{||}^{2}} -4\,p_{2}\cdot r_{\perp
}\,\frac{\partial \Phi _{1}}{\partial r_{\perp }^{2}} -4p_{2}\cdot
r_{||}\,\frac{\partial \Phi _{1}}{\partial r_{||}^{2}}+\{\Phi
_{1},\Phi _{2}\}\approx 0.  \notag \\
&&{}  \label{74}
\end{eqnarray}
The simplest solution is

\begin{equation}
\Phi _{1} = \Phi _{2} = \Phi (\frac{r_{\perp }^{2}}{2}, M),
\label{75}
\end{equation}

\noindent because it implies the following strong satisfaction of the
compatibility condition

\begin{equation}
\{\mathcal{H}_{1}, \mathcal{H}_{2}\} = - 4\, P \cdot r_{\perp }\,
\frac{\partial \Phi ( \frac{r_{\perp }^{2}}{2}, M)}{\partial
r_{\perp }^{2}} = 0.
 \label{76}
\end{equation}

This is the original Droz-Vincent, Todorov, Komar model
\cite{15,16,17}. More general forms of the functions $\Phi_i$ are
possible for which $\{\mathcal{H}_{1}, \mathcal{H}_{2}\} \approx
0$, being proportional to the constraints themselves.

We define the canonical relative momentum by

\begin{equation}
q^{\mu} = \frac{\varepsilon _{2}\, p^{\mu}_{1} - \varepsilon
_{1}\, p^{\mu}_{2}}{M},
  \label{77}
\end{equation}

\noindent with

\begin{equation}
\varepsilon _{1} = \frac{M^{2}+m_{1}^{2}-m_{2}^{2}}{2M}, \qquad
\varepsilon _{2} = \frac{M^{2}+m_{2}^{2}-m_{1}^{2}}{2M}.
 \label{78}
\end{equation}

These constituent particle rest-energies are defined so that

\begin{equation}
\varepsilon _{1}+\varepsilon _{2}= M,\qquad \varepsilon
_{1}-\varepsilon _{2}= \frac{m_{1}^{2}-m_{2}^{2}}{M}.
  \label{79}
\end{equation}
This definition is reinforced by

\begin{eqnarray}
-\frac{p_{1}\cdot P}{M}
&=&\frac{-P^{2}+p_{2}^{2}-p_{1}^{2}}{2M}\approx
\varepsilon _{1},  \notag \\
-\frac{p_{2}\cdot P}{M}
&=&\frac{-P^{2}+p_{1}^{2}-p_{2}^{2}}{2M}\approx \varepsilon _{2},
\label{80}
\end{eqnarray}

\noindent where Eq. (\ref{77}) was used. \ Using $P^{\mu }=p_{1}^{\mu
}+p_{2}^{\mu }$ and Eq.(\ref{77}) gives

\begin{eqnarray}
&&p^{\mu}_{1} = \frac{\varepsilon _{1}P^{\mu}}{M} + q^{\mu},
\qquad
p^{\mu}_{2} = \frac{\varepsilon _{1}P^{\mu}}{M} - q^{\mu}.  \notag \\
&&{}  \label{81}
\end{eqnarray}

In term of these variables the difference of the constraints depends on the
relative energy in the rest frame

\begin{equation}
\mathcal{H}_{1} - \mathcal{H}_{2} = p_{1}^{2} + m_{1}^{2} - p_{2}^{2} -
m_{2}^{2} = 2\, P \cdot q \approx 0,
  \label{82}
\end{equation}

\noindent where we have used

\begin{equation}
\varepsilon _{1}^{2}-m_{1}^{2}=\varepsilon
_{2}^{2}-m_{2}^{2}=\frac{1}{ 4M^{2}
}(M^{4}-2(m_{1}^{2}+m_{2}^{2})M^{2}+(m_{1}^{2}-m_{2}^{2})^{2})\equiv
b^{2}(M).
  \label{83}
\end{equation}

On the other hand, the sum of the two constraints, determines the mass
spectrum of the two-body system. It can be written as

\begin{equation}
q^{2} + \Phi (\frac{r_{\perp }^{2}}{2}, M) - b^{2}(M) \approx 0,
 \label{84}
\end{equation}

\noindent or

\begin{equation}
 {\vec{q}}{}^{2}+\Phi (\frac{\vec{r}^{2}}{2}, M) - b^{2}(M)\approx 0,
 \label{85}
\end{equation}

\noindent in the rest frame (where $q^o \approx 0$ and
$r^2_{\perp} \approx { \vec r}^2$). To get the mass spectrum this
equation has to be solved for $M = \sqrt{P^2}$.

\bigskip

Since we have $\{ x^{\mu}_i, \mathcal{H}_1 \} \not= 0$,
$\{x^{\mu}_i, \mathcal{H}_2 \} \not= 0$, Droz-Vincent covariant
non-canonical positions \cite{13,15} $q^{\mu}_i$ are defined as
the solutions of the two equations $ \{ q^{\mu}_1, \mathcal{H}_2
\} = 0$, $\{ q^{\mu}_2, \mathcal{H}_1 \} = 0$.

\bigskip

If in Eq.(\ref{85}) we consider a $M$-independent potential in the
rest frame we get that

\begin{equation}
{\vec q }{}^{2} = \frac{1}{4M^{2}}\, \Big[M ^{4} - 2\,
(m_{1}^{2}+m_{2}^{2})\, M^{2} + (m_{1}^{2}-m_{2}^{2})^{2}\Big],
\label{86}
\end{equation}

\noindent is modified to

\begin{equation}
{\vec q}{}^{2} + \Phi ({\frac{1}{2}}\, {\vec r }^2)=\frac{1}{4\,
M^{2}}\, \Big[ M^{4}-2(m_{1}^{2}+m_{2}^{2})\, M
^{2}+(m_{1}^{2}-m_{2}^{2})^{2}\Big],
  \label{87}
\end{equation}

\noindent which is the rest-frame form of a covariant two-body constraint
dynamics \cite{23} involving two generalized mass shell constraints of the
form

\begin{equation}
p_{1}^{2}-m_{1}^{2}-\Phi \approx 0,\qquad p_{2}^{2}-m_{2}^{2}-\Phi \approx 0,
 \label{88}
\end{equation}

\noindent with

\begin{equation}
\Phi = 2\, \mu\, V({\frac{1}{2}}\, {\vec r}^2), \qquad \mu =
\frac{m_{1}m_{2} }{m_{1} + m_{2}},
  \label{89}
\end{equation}

\noindent i.e. a form suitable for the non-relativistic limit.

Solving Eq. (\ref{87}) algebraically \ for $M$ and choosing all
positive square roots leads to\footnote{ The choice of
Eq.(\ref{77}) for the relative momentum is the relativistic
generalization of $\vec{q}=(m_{2}\vec{p}_{1}-m_{1}\vec{p}
_{2})/(m_{1}+m_{2})=\mu d\vec{r}/dt$ . \ The alternative choice of
$ q^{\mu} =(p^{\mu}_{1} - p^{\mu}_{2})/2$ would lead to the
constraint $2P\cdot q=m_{1}^{2}-m_{2}^{2}$ in place of
Eq.(\ref{82}). \ However it would lead to the same result,
Eq.(\ref{87}), for $\vec{q}^{2}$ even for unequal mass since the
relative energy is not zero for this choice of $q$ unlike for that
given in Eq.(\ref{77}). \ Hence the expressions in Eqs.(\ref{90})
and (\ref{34}) for the c.m. energy are the same with both choices
of the relative momentum.}

\begin{equation}
M = \sqrt{m_{1}^{2}+{\vec{q}}^{2}+\Phi
({\frac{1}{2}}\,{\vec{r}}^{2})}+\sqrt{
m_{2}^{2}+{\vec{q}}^{2}+\Phi ({\frac{1}{2}}\,{\vec{r}}^{2})}.
\label{90}
\end{equation}
In Section VI the rest-frame instant form of this model is studied in
detail. In particular the form of the generators of the internal Poincare'
group is given.

\section{Inertial and Non-Inertial Rest Frames.}

The inertial rest-frame instant form identifies the 3+1 splitting
of Minkowski space-time corresponding to the inertial rest frame
of every isolated system centered on the inertial observer
associated with the Fokker-Pryce covariant non-canonical external
4-center of inertia. In this instant form the mutual interactions
among the particles of an isolated N-body system are described as
acting on the points of the particle world-lines simultaneous in
the Wigner hyper-planes, leaves of the inertial rest frame.
\medskip

Instead in parametrized Minkowski theories one considers 3+1
splittings of Minkowski space-time corresponding to
\textit{non-inertial frames} centered on arbitrary time-like
observers. If $x^{\mu} = z^{\mu}(\tau ,\vec \sigma )$ describes
the embedding of the associated simultaneity 3-surfaces $
\Sigma_{\tau}$ into Minkowski space-time, so that the metric
induced by the coordinate transformation $x^{\mu} \mapsto \sigma^A
= (\tau ,\vec \sigma )$ is $g_{AB}(\tau ,\vec \sigma ) =
{\frac{{\partial z^{\mu}(\sigma )}}{{\partial \sigma^A}}} \,
\eta_{\mu\nu}\, {\frac{{\partial z^{\nu}(\sigma )}}{{
\partial \sigma^B}}}$ \footnote{
The 4-vectors $z^{\mu}_r(\tau ,\vec \sigma ) = {\frac{{\partial
z^{\mu}(\tau ,\vec \sigma )}}{{\partial \sigma^r}}}$ are tangent
to $\Sigma_{\tau}$. If $ l^{\mu}(\tau ,\vec \sigma )$ is the unit
normal to $\Sigma_{\tau}$ [proportional to
$\epsilon^{\mu}{}_{\alpha\beta\gamma}\, [z^{\alpha}_1\,
z^{\beta}_2\, z^{\gamma}_3](\tau ,\vec \sigma )$], we have $
z^{\mu}_{\tau}(\tau ,\vec \sigma ) = {\frac{{\partial z^{\mu}(\tau
,\vec \sigma )}}{{\partial \tau}}} = N(\tau ,\vec \sigma )\,
l^{\mu}(\tau ,\vec \sigma ) + N^r(\tau ,\vec \sigma )\,
z^{\mu}_r(\tau ,\vec \sigma )$, where $ N(\tau ,\vec \sigma )$ and
$N^r(\tau ,\vec \sigma )$ are the lapse and shift functions,
respectively.}, the basic restrictions on the 3+1 splitting (i.e.
of forming a nice foliation with space-like leaves) are the M$\o
$ller conditions \cite{20}

\begin{eqnarray}
&&\,g_{\tau \tau }(\sigma )>0,  \notag \\
&&{}  \notag \\
&&\,g_{rr}(\sigma )<0,\qquad
\begin{array}{|ll|}
g_{rr}(\sigma ) & g_{rs}(\sigma ) \\
g_{sr}(\sigma ) & g_{ss}(\sigma )
\end{array}
\,>0,\qquad \,det\,[g_{rs}(\sigma )]\,<0,  \notag \\
&&{}  \notag \\
&\Rightarrow &det\,[g_{AB}(\sigma )]\,<0.
  \label{b1}
\end{eqnarray}

\medskip

Moreover, the simultaneity 3-surfaces $\Sigma_{\tau}$ must tend to
space-like hyper-planes at spatial infinity: $z^{\mu}(\tau ,\vec
\sigma ) { \rightarrow}_{|\vec \sigma | \rightarrow \infty}\,\,
x^{\mu}_s(\tau ) + \epsilon^{\mu}_r\, \sigma^r$ and $g_{AB}(\tau
,\vec \sigma ) {\rightarrow} _{|\vec \sigma | \rightarrow
\infty}$\break $\eta_{AB}$, with the $ \epsilon^{\mu}_r$'s being 3
unit space-like 4-vectors tangent to the asymptotic hyper-plane,
whose unit normal is $\epsilon^{\mu}_{\tau}$ [the $
\epsilon^{\mu}_A$ form an asymptotic cotetrad, $\epsilon^{\mu}_A\,
\eta^{AB}\, \epsilon^{\nu}_B = \eta^{\mu\nu}$].

\medskip

As shown in Refs.\cite{20,36}, Eqs.(\ref{b1}) forbid rigid
rotations: \textit{only differential rotations} are allowed and
the simplest example is given by those 3+1 splittings whose
simultaneity 3-surfaces are hyper-planes with rotating
3-coordinates described by the embeddings ($\sigma =|\vec{
\sigma}|$)

\begin{eqnarray}
&&z^{\mu }(\tau ,\vec{\sigma})=x^{\mu }(\tau )+\epsilon _{r}^{\mu
}\,R^{r}{}_{s}(\tau ,\sigma )\,\sigma ^{s},  \notag \\
&&{}  \notag \\
&&R^{r}{}_{s}(\tau ,\sigma ){\rightarrow }_{\sigma \rightarrow
\infty }\delta _{s}^{r},\qquad \partial _{A}\,R^{r}{}_{s}(\tau
,\sigma )\,{
\rightarrow }_{\sigma \rightarrow \infty }\,0,  \notag \\
&&{}  \notag \\
&&R(\tau ,\sigma )=\tilde{R}(\beta _{a}(\tau ,\sigma )),\qquad \beta
_{a}(\tau ,\sigma )=F(\sigma )\,{\tilde{\beta}}_{a}(\tau ),\quad a=1,2,3,
\notag \\
&&{}  \notag \\
&&{\frac{{dF(\sigma )}}{{d\sigma }}}\not=0,\qquad 0<F(\sigma
)<{\frac{1}{{ A\,\sigma }}}.
  \label{b2}
\end{eqnarray}

Each $F(\sigma )$ satisfying the restrictions of the last line, coming from
Eqs.(\ref{b1}), gives rise to a \textit{global differentially rotating
non-inertial frame}. \bigskip

As shown in Refs.\cite{2}, given the Lagrangian of every isolated
system, one makes the coupling to an external gravitational field
and then replaces the external metric with the $g_{AB}(\tau
,\vec{\sigma})$ associated to a M$\o$ller-admissible 3+1
splitting. The resulting action principle $S=\int d\tau
d^{3}\sigma \,\mathcal{L}(matter,g_{AB}(\tau ,\vec{\sigma}))$
depends upon the system and the embedding $z^{\mu }(\tau
,\vec{\sigma})$ and is invariant under frame-preserving
diffeomorphisms: $\tau \mapsto \tau ^{^{\prime }}(\tau
,\vec{\sigma})$, $\sigma ^{r}\mapsto \sigma ^{{^{\prime }}
\,r}(\vec{\sigma})$. This special-relativistic general covariance
implies the vanishing of the canonical Hamiltonian and the
following 4 first class constraints

\begin{eqnarray}
\mathcal{H}_{\mu }(\tau ,\vec{\sigma}) &=&\rho _{\mu }(\tau
,\vec{\sigma} )-\epsilon \,l_{\mu }(\tau
,\vec{\sigma})\,\mathcal{M}(\tau ,\vec{\sigma} )-\epsilon
\,z_{r\mu }(\tau ,\vec{\sigma})\,h^{rs}(\tau ,\vec{\sigma})\,
\mathcal{M}_{s}(\tau ,\vec{\sigma})\approx 0,  \notag \\
&&\{\mathcal{H}_{\mu }(\tau ,\vec{\sigma}),\mathcal{H}_{\nu }(\tau
,{\vec{ \sigma}}^{^{\prime }})\}=0,
  \label{b3}
\end{eqnarray}

\noindent where $\rho _{\mu }(\tau ,\vec{\sigma})$ is the momentum
conjugate to $z^{\mu }(\tau ,\vec{\sigma})$ and
$[\sum_{u}\,h^{ru}\,g_{us}](\tau ,\vec{ \sigma})=\delta _{s}^{r}$.
$\mathcal{M}(\tau ,\vec{\sigma})=T_{\tau \tau }(\tau
,\vec{\sigma})$ and $\mathcal{M}_{r}(\tau ,\vec{\sigma})=T_{\tau
r}(\tau ,\vec{\sigma})$ are the energy- and momentum- densities of
the isolated system in $\Sigma _{\tau }$-adapted coordinates
\footnote{For N free particles we have $\mathcal{M}(\tau
,\vec{\sigma})=\sum_{i=1}^{N}\,\delta
^{3}(\vec{\sigma}-{\vec{\eta}}_{i}(\tau
))\,\sqrt{m_{i}^{2}+h^{rs}(\tau , \vec{\sigma})\,\kappa _{ir}(\tau
)\,\kappa _{is}(\tau )}$, $\mathcal{M} _{r}(\tau
,\vec{\sigma})=\sum_{i=1}^{N}\,\delta
^{3}(\vec{\sigma}-{\vec{\eta} }_{i}(\tau ))\,\kappa _{ir}(\tau
)$.}.

Since the matter variables have only $\Sigma_{\tau}$-adapted Lorentz-scalar
indices, the 10 constant of the motion corresponding to the generators of
the external Poincare' algebra are

\begin{eqnarray}
P^{\mu } &=&\int d^{3}\sigma \,\rho ^{\mu }(\tau ,\vec{\sigma}),  \notag \\
J^{\mu \nu } &=&\int d^{3}\sigma \,[z^{\mu }\,\rho ^{\nu }-z^{\nu }\,\rho
^{\mu }](\tau ,\vec{\sigma}).
  \label{b4}
\end{eqnarray}

\medskip

The Hamiltonian gauge transformations generated by the constraints
(\ref{b3} ) change the form and the coordinatization of the
simultaneity 3-surfaces $ \Sigma _{\tau }$ (on each of these
surfaces all the clocks are synchronized): as a consequence the
embeddings $z^{\mu }(\tau ,\vec{\sigma })$ are \textit{gauge
variables}, so that the choice of the non-inertial frame and in
particular of the convention for the synchronization of distant
clocks \cite{20} is a \textit{gauge choice} in this framework. All
the inertial and non-inertial frames compatible with the M$\o
$ller conditions ( \ref{b1}) are \textit{gauge equivalent} for the
description of the dynamics of isolated systems.

The inertial rest-frame instant form of Section III is associated to the
special gauge $z^{\mu }(\tau ,\vec{\sigma})=x_{s}^{\mu }(\tau )+\epsilon
_{r}^{\mu }(u(P))\,\sigma ^{r}$, $x_{s}^{\mu }(\tau )=Y_{s}^{\mu }(\tau
)=u^{\mu }(P)\,\tau $, selecting the inertial rest frame of the isolated
system centered on the Fokker-Pryce 4-center of inertia and having as
instantaneous 3-spaces the Wigner hyper-planes, whose internal 3-vectors
transform as Wigner spin-1 3-vectors.

\bigskip

Another particularly interesting family of 3+1 splittings of Minkowski
space-time is defined by the embeddings

\begin{eqnarray}
z^{\mu }(\tau ,\vec{\sigma}) &=&Y_{s}^{\mu }(\tau )+F^{\mu }(\tau
,\vec{ \sigma})=u^{\mu }(P)\,\tau +F^{\mu }(\tau
,\vec{\sigma}),\qquad F^{\mu
}(\tau ,\vec{0})=0,  \notag \\
&{\rightarrow }_{\sigma \rightarrow \infty }& u^{\mu }(P)\,\tau
+\epsilon _{r}^{\mu }(u(P))\,\sigma ^{r},
  \label{b5}
\end{eqnarray}

\noindent with $F^{\mu }(\tau ,\vec{\sigma})$ satisfying Eqs.(\ref{b1}).
\medskip

In this family the simultaneity 3-surfaces $\Sigma_{\tau}$ tend to
Wigner hyper-planes at spatial infinity, where they are orthogonal
to a 4-vector $P^{\mu} = M\, u^{\mu}(P)$, where $M =
\sqrt{P^2_{sys}}$ is the invariant mass of the isolated system. As
a consequence, there are asymptotic inertial observers with the
world-lines parallel to that of the Fokker-Pryce 4-center of
inertia, namely there are the \textit{rest-frame conditions} $ p_r
= \epsilon^{\mu}_r(u(P))\, P_{\mu} = 0$, so that the embeddings
(\ref{b5} ) define \textit{global M$\o $ller-admissible
non-inertial rest frames} \footnote{ The only ones existing in
tetrad gravity, due to the equivalence principle, in globally
hyperbolic asymptotically flat space-times without
super-translations as shown in Ref.\cite{23} and its
bibliography.}. \medskip

In these non-inertial rest frames the external Poincare'
generators \footnote{ They are fixed by boundary conditions giving
the value of constants of the motion corresponding to asymptotic
symmetries of these 3+1 splittings.} are given by Eqs.(\ref{b4})
evaluated on the embeddings (\ref{b5}) and taking into account the
constraints (\ref{b3}). The internal Poincare' generators $M$,
$p^r \approx 0$, $j^r$, $k^r$ can be identified from the
energy-momentum tensor of the isolated system restricted to the
embedding (\ref{b5}). While $M$ is the effective Hamiltonian for
the motion of the relative variables, $k^r \approx 0$ is the
natural gauge fixing for the rest-frame conditions $p_r \approx 0$
.\medskip

Since we are in non-inertial rest frames, inside the internal
energy- and boost- densities there are the inertial potential
sources of the relativistic inertial forces \cite{36}: they are
contained in the spatial components of the metric $g_{rs}(\tau
,\vec{\sigma})$ associated with the embeddings (\ref{b5}) and
describe the {\it appearances} of phenomena in non-inertial
frames.
\medskip

For N-body systems with a-a-a-d interactions each non-inertial
rest frame ( \ref{b5}) describes the same world-lines with
different correlations among them, determined by their
intersection with the simultaneity 3-surfaces $ \Sigma_{\tau}$,
and with the extra inertial forces. The difficulty in obtaining
these non-inertial descriptions is the necessity to have an
explicit Lagrangian with interactions to be used as a starting
point. For instance they could be studied for the systems of Refs.
\cite{27,28} containing the electro-magnetic field.

\vfill\eject

\end{document}